# General solution of the Schrödinger equation with potential field quantization

**Hasan Huseyin Erbil**

Ege University, Science Faculty, Izmir/Turkey

hhuseyin.erbil@gmail.com, huseyin.erbil41@ttmail.com

***Abstract** – It has been found a simple procedure for the general solution of the time-independent Schrödinger equation (SE) with the help of quantization of potential area in one dimension without making any approximation. Energy values are not dependent on wave functions, so to find the energy values; it is enough to find the classic turning points of the potential function. Two different solutions were obtained, namely, symmetric and antisymmetric at bound states. These normalized wave functions are always periodic. It is enough to take the integral of the square root of the potential energy function to find the normalized wave functions. If these calculations cannot be made analytically, it should then be performed by numerical methods. SE has been solved for a particle in many one-dimension and the spherical symmetric central potential well as examples. It has been found their energies and normalized wave functions as examples. These solutions were also applied to the theories of scattering and alpha decay. The results obtained with the experimental values were compared with the calculated values. One has been seen to be very fit.*

**1. Introduction**

Mechanical total energy (E) is the sum of kinetic energy (T) and potential energy (U). That is, $E = T + U$. If $E = U$ then $T = 0$. If $E > U$ then $T > 0$ ; If $E < U$ then $T < 0$ (it is not possible, classically). In quantum mechanics, the total energy is equal to the eigenvalue of the total energy operator $\hat{H}$ (Hamiltonian). One dimension Hamiltonian is: $\hat{H} = -\frac{\hbar^2}{2m}\frac{d^2}{dx^2} + U(x)$. This operator is hermitic and its eigenvalues are real numbers. The equation of eigenvalues of this Hamiltonian is given as follows (time-independent Schrödinger Equation, SE):

$$\hat{H}\,\psi(x) = \left[-\frac{\hbar^2}{2m}\frac{d^2}{dx^2} + U(x)\right]\psi(x) = E\,\psi(x) \qquad (1.1)$$

The eigenvalue E can have two values, $-E$ and $+E$, $(E > 0)$. So from equation (1.1), the following differential equation is obtained:

$$\frac{d^2\psi(x)}{dx^2} + [k^2 - m_1^2\,U(x)]\,\psi(x) = 0 \qquad (1.2)$$

Here, $k^2 = -m_1^2\,E$ for $-E$ and $k^2 = m_1^2\,E$ for $+E$; $m_1^2 = 2m/\hbar^2$ . (1.3)

Although we succeed in solving the time-independent SE for some quantum mechanical problems in one dimension, an exact solution is not possible in complicated situations, and we must then resort to approximation methods. For the calculation of stationary states and energy eigenvalues, these include perturbation theory, the variation method, the Method of Nikiforov-Uvarov [1-4], the Supersymmetric Quantum Mechanics, the Supersymmetric WKB and the WKB approximations [5, 6], Perturbation theory, etc. Perturbation theory is applicable if the Hamiltonian differs from an exactly solvable part by a small amount. The variation method [18, 37] is appropriate for the calculation of the ground state energy if one has a qualitative idea of the form of the wave function. The WKB method [13, 15] is applicable in the nearly classical limit. Until now, the general solution of the differential equation given by (1.2) has not been made exactly. This problem is a very challenging problem for theoretical physicists.

In this study we achieved a simple procedure for the exact general solution of the time-independent SE in one dimension without making any approximation. We have applied this simple procedure to various quantum mechanical problems: one-dimension potentials, spherically symmetric potentials, tunneling effect, scattering theory and some examples.

## 2. Solution of time-independent Schrödinger equation in one dimension

Let us rewrite the time-independent Schrödinger Equation (SE) in one dimension (1.2):

$$\frac{d^2\psi(x)}{dx^2} + [k^2 - m_1^2\, U(x)]\, \psi(x) = 0 \tag{2.1}$$

In (2.1), E and $U(x)$ are respectively the total and effective potential energies of a particle of mass m. Two kinds of solutions to the SE correspond precisely to bound and scattering (unbound) states. When $E < U(-\infty)$ and $U(+\infty)$, solutions correspond to bound states; when $E > U(-\infty)$ or $U(+\infty)$, solutions correspond to scattering states [6]. In real life, many potential functions go to zero at infinity, in which case the criterion is simplified even further: when $E < 0$ bound state occurs, when $E > 0$ scattering state occurs. In this section, we shall explore potentials that give rise to both kinds of states. We shall resolve this equation [equation (2.1)] by two steps:

*2.1. First Step*

By inspiration of the theorem given in refs [8, 9], we consider the integral function:

$$S(x) = \int U(x)dx. \tag{2.2}$$

If $f(x) = S(x)$ is taken in this theorem, we get: $F(\varepsilon) = \int_{-\infty}^{+\infty} S(x)\, y(x - x_0, \varepsilon)\, dx \tag{2.3}$

From this function for $\varepsilon = 0$, it is obtained,

$$\lim_{\varepsilon \to 0} F(\varepsilon) = F(0) = \lim_{\varepsilon \to 0} \int_{-\infty}^{+\infty} S(x)\delta(x - x_0)dx = S(x_0) = S \tag{2.4}$$

Now, let's take the potential and wave function, respectively, as follows:

$$S = \int_{x_1}^{x_2} U(x)\, dx; \quad U(x) = S(x_0)\, \delta(x - x_0) = S\, \delta(x - x_0) \text{ and } \psi(x) = F(x) \tag{2.5}$$

$\delta(x-x_0)$ is Dirac function. $x_1$ and $x_2$ are the roots of the equation $E = U(x)$, that is the apsis of the classical turning point of the potential function. If we take $x_0 = (x_1 + x_2)/2$ and $d = x_2 - x_1$, $(x_2 > x_1)$, then we find $x_1 = x_0 - d/2$, $x_2 = x_0 + d/2$. (See figure 1). If we take these functions, the SE (2.1) becomes as follows:

$$\frac{d^2F(x)}{dx^2} + k^2 F(x) = m_1^2 S\, \delta(x - x_0)\, F(x) \tag{2.6}$$

To evaluate the behavior of $F(x)$ at $x = x_0$, let us integrate the equation (2.6) over the interval $[x_1, x_2] = [x_0 - d/2, x_0 + d/2]$ and also consider the limit $d \to 0$, then we obtain,

$$F'(x_0 + d/2) - F'(x_0 - d/2) = m_1^2 S\, F(x_0) \tag{2.7}$$

This equation (2.7) shows that the derivation of F(x) is not continuous at the $x = x_0$ point [6, 9], whereas the wave functions F(x) should be continuous at the same point.

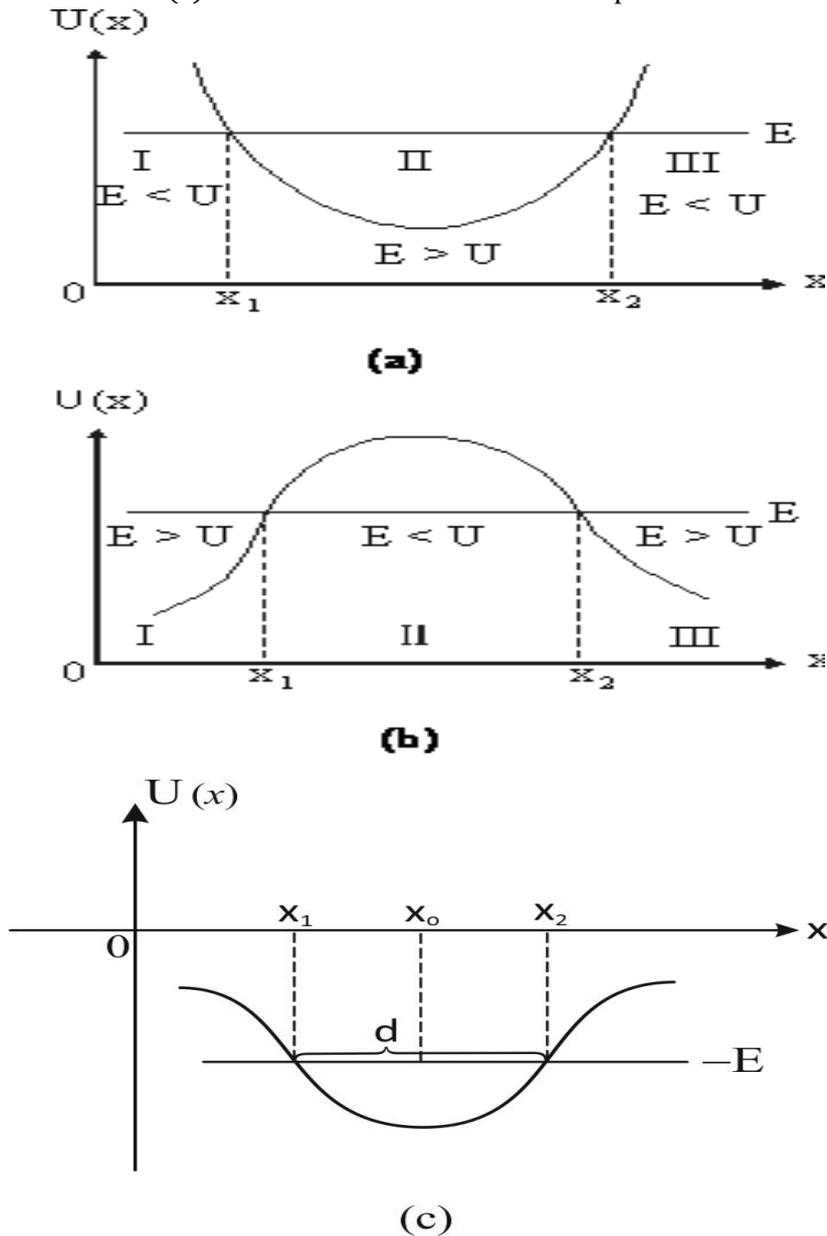

**Figure 1.** Regions relevant to a particle of energy E moving in a one dimensional potential field U(x). **(a)** In the domains I and III, $E < U(x)$, (unbound state); In the domain II, $E > U(x)$, (bound state). **(b)** In the domains I and III, $E > U(x)$, In the domain II, $E < U(x)$, (unbound state). The roots of the equation $E = U(x)$ are turning points of the corresponding classical motion. **(c)** Regions for $U(x) < 0$ and $E < 0$

To solve the differential equation (2.6), we can perform the transformation of Fourier of the equation and then we obtain the following function:

$$F(x) = A\, e^{-k\,|(x-x_0)|}\ ,\ \left[A = \frac{a^2}{k}\sqrt{\frac{\pi}{2}}\ ,\ a^2 = -\frac{2m}{\hbar^2\sqrt{2\pi}}\, S\, F(x_0)\right] \tag{2.8}$$

From the equation (2.8), we get:

$$F(x) = A\, e^{k(x-x_0)} \text{ for } x < x_0 \text{ and } F(x) = A\, e^{-k(x-x_0)} \text{ for } x > x_0 \tag{2.9}$$

Substituting the functions (2.9) into the equation (2.7) and taking the limit $d \to 0$, it is obtained:

$$k = \frac{m}{\hbar^2} S \text{ or } E = \pm \frac{m}{2\hbar^2} S^2 \tag{2.10}$$

It can also be obtained the same values (2.10) from the equation (2.8). To find the constant A, the function $F(x)$ can be normalized to 1:

$$\int_{-\infty}^{x_0} A\, A^*\, e^{2k(x-x_0)} dx + \int_{x_0}^{+\infty} A\, A^*\, e^{-2k(x-x_0)} dx = 1$$

From this equation, we obtain the coefficient of normalization as:

$$|A| = \sqrt{k} = \sqrt{m\,S}/\hbar \tag{2.11}$$

From (2.9), by the linear combinations of these functions, we have also the following functions:

$$F(x) = A\, e^{k(x-x_0)} + B\, e^{-k(x-x_0)} \tag{2.12a}$$

$$F(x) = \frac{1}{2} A\left[e^{k(x-x_0)} + e^{-k(x-x_0)}\right] = A \cosh[k(x - x_0)] \tag{2.12b}$$

$$F(x) = \frac{1}{2} A\left[e^{k(x-x_0)} - e^{-k(x-x_0)}\right] = A \sinh[k(x - x_0)] \tag{2.12c}$$

## 2.2. Second Step

Let us assume the wave function $\psi(x)$ to be $\boldsymbol{\psi(x) = F(x)\, e^{i\,G(x)}}$, (we assume that $G(x)$ is a real function). If we substitute this function into (2.1), we get:

$$F''(x) - F(x)\,G'^{2}(x) - k^2 F(x) - m_1^2\, U(x)F(x) + i\,[2\,F'(x)G'(x) + F(x)G''(x)] = 0 \tag{2.13}$$

From the real and imaginary parts of the equation (2.13), we can have the following two equations:

$$F''(x) - F(x)G'^{2}(x) - k^2 F(x) - m_1^2\, U(x)F(x) = 0 \tag{2.14}$$

$$2\,F'(x)G'(x) + F(x)G''(x) = 0 \tag{2.15}$$

From the equation (2.14), the following equations are obtained:

For $F(x) = A\, e^{k(x-x_0)}$;  $F(x)\left[m_1^2 U(x) + G'^{2}(x)\right] = 0 \tag{2.16a}$

For $F(x) = A\, e^{-k(x-x_0)}$;  $F(x)\left[m_1^2 U(x) + G'^{2}(x)\right] = 0 \tag{2.16b}$

For $F(x) = A\, e^{k(x-x_0)} + B\, e^{-k(x-x_0)}$;  $F(x)\left[m_1^2 U(x) + G'^{2}(x)\right] = 0 \tag{2.16c}$

For $F(x) = A \cosh[k(x - x_0)]$;  $F(x)\left[m_1^2 U(x) + G'^{2}(x)\right] = 0 \tag{2.16d}$

For $F(x) = A \sinh[k(x - x_0)]$;  $F(x)\left[m_1^2 U(x) + G'^{2}(x)\right] = 0 \tag{2.16e}$

From the equation (2.16), the following two equations are obtained:

$$m_1^2 U(x) + G'^{2}(x) = 0 \tag{2.17a}$$

$$F(x) = 0 \tag{2.17b}$$

From the equation (2.17a) the function $G(x)$ is obtained as follows:

$$G(x) = \pm\, m_1 \int \sqrt{-U(x)}\, dx = \pm\, i\, m_1 \int \sqrt{U(x)}\, dx = \pm i\, Q(x) \tag{2.18a}$$

$$Q(x) = m_1 \int \sqrt{U(x)}\, dx \tag{2.18b}$$

Thus, the wave function $\psi(x)$ has been written as follows:

$$\psi(x) = F(x)\, e^{\pm i\, G(x)} \text{ or } \psi(x) = F(x - x_0)\, e^{\pm i\, G(x-x_0)} \tag{2.19}$$

The functions (2.19) can also be written as follows:

$$\boldsymbol{\psi(x) = F(x)\left[A\, e^{i\,G(x)} + B\, e^{-i\,G(x)}\right]} \tag{2.20a}$$

$$\boldsymbol{\psi(x) = F(x - x_0)\left[A\, e^{i\,G(x-x_0)} + B\, e^{-i\,G(x-x_0)}\right]} \tag{2.20b}$$

In the functions (2.20): **(a)** For $E > U(x)$ ; $k = m_1 \sqrt{-E}$, $G(x) = m_1 \int \sqrt{-U(x)}\, dx$ ;

**(b)** For $E < U(x)$; ; $k = m_1 \sqrt{E}$, $G(x) = m_1 \int \sqrt{U(x)}\, dx \qquad (2.20c)$

As shown in figure 1a, the total energy is equal to potential energy at the points $x_1 = x_0 - d/2$ and $x_2 = x_0 + d/2$. So kinetic energy is zero at these points. Now, let us examine the behavior of the function in the interval $[x_0 - d/2, x_0 + d/2]$. Let us consider the equation (2.17b), that is:

$$F(x) = 0 \tag{2.17b}$$

Let us assume that the roots of this equation are: $x_1 = x_0 - d/2$ and $x_2 = x_0 + d/2$. So we can write: $F(x_1) = F(x_0 - d/2) = 0$ ; $F(x_2) = F(x_0 + d/2) = 0$ \hfill (2.21)

From equations (2.21), the following expressions can be obtained:

**(a)** For $F(x) = A\, e^{k(x-x_0)} + B\, e^{-k(x-x_0)}$; $k = m_1\sqrt{-E} = i\, m_1\sqrt{E} = i\, K$, $[K = m_1\sqrt{E} > 0]$

$a_{11}A + a_{12}B = 0;\quad a_{21}A + a_{22}B = 0;$

$a_{11} = e^{-i\,d\,K/2}\,;\ a_{12} = e^{i\,d\,K/2};\ a_{21} = e^{i\,d\,K/2};\ a_{22} = e^{-i\,d\,K/2}$

Since A and B have non-zero values, the determinant of coefficients in this system of equations must be zero. Thus the following equation can be written:

$\det\begin{vmatrix} a_{11} & a_{12} \\ a_{21} & a_{22} \end{vmatrix} = -2\, i\, \sin(d\,K) = 0.$

Here let us take $d\,K = q$. From this equation, we have

$$\sin(d\,K) = \sin(q) = 0 \ \to\ q = n\,\pi,\ [n = 1, 2, 3, 4, \ldots \text{integer}] \tag{2.22}$$

So, we obtain the **quantization condition of energy** as follows:

$$\mathbf{K\,d = \sqrt{\tfrac{2\,m}{\hbar^2}\,|E|}\ d = q,\ [q = n\,\pi,\ (n = 1, 2, 3, \ldots\ \text{integer numbers})]} \tag{2.23}$$

Here the coefficient B is obtained as $B = -A$. Thus the function $F(x)$ is written as follows:

$F(x) = A\bigl[e^{i\,K\,(x-x_0)} - e^{-i\,K\,(x-x_0)}\bigr] = 2\,i\,A\sin[K(x-x_0)] = B\sin[K(x-x_0)]$

The coefficient B in this function is found by normalizing of function in the interval $[x_1, x_2]$. One is found: $|B| = \sqrt{2/d}$. Thus the functions $F(x)$ and $\psi(x)$ are written as follows:

$$\mathbf{F(x) = \sqrt{\tfrac{2}{d}}\ \sin[K(x-x_0)]\ ;\ \ \psi(x) = \sqrt{\tfrac{2}{d}}\ \sin[K(x-x_0)]\ e^{iG(x-x_0)}} \tag{2.24}$$

**(b)** For $F(x) = A\cosh[k(x-x_0)]$; $k = i\,K$

$F(x_1) = F(x_0 - d/2) = 0$ ; $F(x_2) = F(x_0 + d/2) = 0$ ;

$F(x_1) = F(x_2) = A\cosh(d\,k/2) = A\cos(d\,K/2) = 0$

$d\,K/2 = q/2 = (2\,n - 1)\,\pi/2\ \to q = (2\,n - 1)\pi,\ [n = 1, 2, 3, \ldots \text{integer}]$

So, we obtain the quantization condition of energy as follows (symmetric case):

$$\mathbf{K\,d = \sqrt{\tfrac{2\,m}{\hbar^2}\,|E|}\ d = q,\ [q = (2n - 1)\,\pi,\ (n = 1, 2, 3, \ldots\ \text{integer numbers})]} \tag{2.25}$$

The coefficient A in this function is found by normalizing the function in the interval $[x_1, x_2]$. That is:

$|A| = \sqrt{2/d}$ . Thus the functions $F(x)$ and $\psi(x)$ are written as follows:

$$F(x) = \sqrt{\frac{2}{d}} \cos[K(x - x_0)] \; ; \; \psi(x) = \sqrt{\frac{2}{d}} \cos[K(x - x_0)] \, e^{i\,G(x-x_0)} \tag{2.26}$$

(c) For $F(x) = A \sinh[k(x - x_0)]$; $k = i K$ ;
$F(x_1) = F(x_0 - d/2) = 0$ ; $F(x_2) = F(x_0 + d/2) = 0$ ;
$F(x_1) = -A \sinh(d\,k/2) = -i A \sin(d\,K/2) = 0$
$F(x_2) = A \sinh(d\,k/2) = i A \sin(d\,K/2) = 0$
$d\,K/2 = q/2 = n\pi \; \to q = 2n\pi$, $[n = 1, 2, 3, \ldots$ integer numbers$]$

So, we obtain the quantization condition of energy as follows (antisymmetric case):

$$\mathbf{K\,d} = \sqrt{\frac{2m}{\hbar^2} |E|}\; \mathbf{d} = \mathbf{q},\; [\mathbf{q = 2n\pi,\; (n = 1, 2, 3, \ldots\; integer\; numbers)}] \tag{2.27}$$

The coefficient $A$ in this function is found by normalizing the function in the range $[x_1, x_2]$. That is: $|A| = \sqrt{2/d}$ . Thus the functions $F(x)$ and $\psi(x)$ are written as follows:

$$F(x) = \sqrt{\frac{2}{d}} \sin[K(x - x_0)] \; ; \; \psi(x) = \sqrt{\frac{2}{d}} \sin[K(x - x_0)] \, e^{i\,G(x-x_0)} \tag{2.28}$$

It is possible to combine the equations (2.25) and (2.27) in one equation as follows (general case):

$$\mathbf{K\,d} = \sqrt{\frac{2m}{\hbar^2} |E|}\; \mathbf{d} = \mathbf{q},\; [\mathbf{q = n\pi,\; (n = 1, 2, 3, \ldots\; integer\; numbers)}] \tag{2.23}$$

Now let us write the kinetic energy of the particle as follows: $T = \frac{p^2}{2m} = E - U(x)$. By integrating this equation from $x_1$ to $x_2$ and using the equation (2.5), the following equation is obtained:

$\int_{x_1}^{x_2} T\, dx = \int_{x_1}^{x_2} \frac{p^2}{2m}\, dx = \int_{x_1}^{x_2}[E - U(x)]dx = \int_{x_1}^{x_2} E\, dx - \int_{x_1}^{x_2} U(x)\, dx = S_k$ and the following
equation is written: $S_k = E(x_2 - x_1) - S = E\,d - S$ \hfill (2.29)

From the figure 1, we can observe that:
(a) For the case $E > U(x)$, the kinetic energy is positive and $[E(x_2 - x_1) - S] > 0$ (bound state).
(b) For the case $E < U(x)$, the kinetic energy is imaginary and $[E(x_2 - x_1) - S] < 0$ (unbound state).
(c) For the case $E = U(x)$, the kinetic energy is zero and $[E(x_2 - x_1) - S] = 0$ (ground state).

In addition, for the bound states, in the interval $[x_1, x_2]$, the kinetic energy is positive, outside this interval, the kinetic energy is imaginary. The minimum point of the potential corresponds to ground state. At ground state, the kinetic energy is zero, namely $T = 0$ or $S_k = 0$. Thus, at the minimum point of the potential, we can write that: $E_0(x_2 - x_1) - S = 0$ or $S = E_0(x_2 - x_1) = E_0 d$. By substituting this value of S into the equation (2.10), we get the ground state energy expression as follows:

$$E_0 = -\frac{m}{2\,\hbar^2} S^2 = -\frac{m}{2\,\hbar^2} E_0^2 d^2 \quad \to \quad E_0 = -\frac{2\,\hbar^2}{m\,d^2} \tag{2.30}$$

Here $E_0$ represents the ground state energy. The negative sign indicates that the state is bound and it can be omitted for the positive energies in the calculations.

### 3. Boundary conditions
Let us divide the potential field into three domains as shown in figure 1 and represent the functions $\psi_1(x)$, $\psi_2(x)$ and $\psi_3(x)$ in each domain. The wave functions and their derivatives should be continuous. Because of these conditions, the above functions must satisfy the following conditions:

$$\psi_1(x_1) = \psi_2(x_1) \,;\, \psi_1'(x_1) = \psi_2'(x_1) \,;\, \psi_2(x_2) = \psi_3(x_2) \,,\, \psi_2'(x_2) = \psi_3'(x_2)$$
$$\lim_{x \to -\infty} \psi_1(x) \to 0 \quad \text{and} \quad \lim_{x \to +\infty} \psi_3(x) \to 0 \tag{3.1}$$

The normalization of the bound state function requires that the functions vanish at infinity. With these boundary and normalization conditions of the wave functions, we can find the integral constants, $A$, $B$ and the energy $E$. As it will be also seen above, in the bound states, we do not need the solutions of the SE. It is sufficient to know only the classical turning points, $x_1$ and $x_2$. nevertheless, the quantum energy values found above can also be found with the help of boundary conditions. Now, in bound states, let us apply the conditions (3.1) to the following functions:

$$A\,e^{k(x-x_0)} + B\,e^{-k(x-x_0)} \,;\, A\cosh[k(x-x_0)] \,;\, B\sinh[k(x-x_0)].$$

**(a)** For function $A\,e^{k(x-x_0)} + B\,e^{-k(x-x_0)}$

According to figure 1a, in the domain I, $E < U(x)$, in the domain II, $E > U(x)$, in the domain III, $E < U(x)$. According to the equations (2.20), the corresponding wave functions are written as follows:

$$\psi_1(x) = A_1 e^{i\,k(x-x_0)} e^{-Q(x-x_0)} \tag{3.2a}$$
$$\psi_2(x) = [A_2 e^{k(x-x_0)} + B_2 e^{-k(x-x_0)}]\, e^{i\,Q(x-x_0)} \tag{3.2b}$$
$$\psi_3(x) = A_3 e^{-i\,k(x-x_0)} e^{-Q(x-x_0)} \tag{3.2c}$$

Here, $Q(x-x_0) = m_1 \int \sqrt{-U(x-x_0)}\, dx$, $k = m_1\sqrt{-E} = i\,m_1\sqrt{E}$

Boundary conditions: $\psi_1(x_1) = \psi_2(x_1)$, $\psi_2(x_2) = \psi_3(x_2)$, $\psi_1'(x_1) = \psi_2'(x_1)$, $\psi_2'(x_2) = \psi_3'(x_2)$
$$\lim_{x \to -\infty} \psi_1(x) \to 0 \quad \text{and} \quad \lim_{x \to +\infty} \psi_3(x) \to 0$$

From these conditions, four linear equations are obtained as follows:

$$\begin{bmatrix} a_{11} & a_{12} & a_{13} & a_{14} \\ a_{21} & a_{22} & a_{23} & a_2 \\ a_{31} & a_{32} & a_{33} & a_{34} \\ a_{41} & a_{42} & a_{43} & a_{44} \end{bmatrix} \begin{bmatrix} A_1 \\ A_2 \\ B_2 \\ A_3 \end{bmatrix} = 0 \tag{3.3a}$$

In order this system of equations to have a solution different from zero, the determinant of coefficients should vanish, namely;

$$\det \begin{vmatrix} a_{11} & a_{12} & a_{13} & a_{14} \\ a_{21} & a_{22} & a_{23} & a_2 \\ a_{31} & a_{32} & a_{33} & a_{34} \\ a_{41} & a_{42} & a_{43} & a_{44} \end{vmatrix} = 0 \tag{3.3b}$$

If it is taken: $x_1 = x_0 - d/2$, $x_2 = x_0 + d/2$, $Q'(-d/2) = Q'(d/2) = k$, $k = i K$, $d K = q$, from (3.3), it is obtained:

$$e^{[-i q-(1-i)Q(-d/2)+(1+i)Q(d/2)]} \left[-1 + e^{2 i q}\right] = 0 \tag{3.4}$$

In order the equality (3.4) to be realized, it has to be $q = n \pi$, ($n = 1, 2, 3, \ldots$ integer numbers). So, we obtain the quantization condition of energy as follows:

$$\mathbf{K\, d = \sqrt{\frac{2\,m}{\hbar^2}\,|E|}\;d = q,\;[q = n\,\pi,\;(n = 1, 2, 3, \ldots \text{integer numbers})]} \tag{3.5}$$

**(b)** For function $\cosh[k\,(x - x_0)]$

According to the figure 1a, in the domain I, $E < U(x)$, in the domain II, $E > U(x)$, in the domain III, $E < U(x)$. According to the equations (2.20), the corresponding wave functions are written as follows:

$\psi_1(x) = A_1 \cosh[i\,k\,(x - x_0)]\,e^{i\,i\,G(x-x_0)}$ ; $\psi_2(x) = A_2 \cosh[k\,(x - x_0)]\,e^{i\,G(x-x_0)}$
$\psi_3(x) = A_3 \cosh[-i\,k\,(x - x_0)]\,e^{i\,i\,G(x-x_0)}$ ; $x_1 = x_0 - d/2$ and $x_2 = x_0 + d/2$

According to the conditions (3.1) it can be written:

$\psi_1(x_1) = \psi_2(x_1)$ or $\psi_1(x_1) - \psi_2(x_1) = 0$ ; $\psi_2(x_2) = \psi_3(x_2)$ or $\psi_2(x_2) - \psi_3(x_2) = 0$ (3.6)

From (3.6) two linear equations are obtained as follows:

$$a_{11} A_1 + a_{12} A_3 = 0; \quad a_{21} A_1 + a_{22} A_3 = 0 \tag{3.7}$$

Here, $a_{11} = e^{-G(-d/2)} \cos(d\,k/2)$, $a_{12} = 0$; $a_{21} = 0$, $a_{22} = -e^{G(d/2)} \cos(d\,k/2)$ (3.8)

In order this system of equations to have a solution different from zero, the determinant of coefficients should vanish, namely,

$$\det \begin{vmatrix} a_{11} & a_{12} \\ a_{21} & a_{22} \end{vmatrix} = 0 \;\rightarrow\; -\frac{1}{2} e^{-G(-d/2)+G(d/2)} [1 + \cos(d\,k)] = 0 \tag{3.9}$$

In equation (3.9), $-\frac{1}{2} e^{-G(-d/2)+G(d/2)} \neq 0$, $k = m_1 \sqrt{-E} = i\,m_1 \sqrt{E} = i K$, ($E > 0$, $K = m_1 \sqrt{E}$)

So, $1 + \cos(d\,k) = 1 + \cos(i d\,K) = 1 + \cos(d\,K) = 0$, from this last equation, we have that:

$d\,K = q = n \pi$, ($n = 1, 3, 5, 7, \ldots$ impair integer). This is also written as follows:

$$\mathbf{d\,K = q = (2\,n - 1)\pi,\;(n = 1, 2, 3, \ldots \text{integer})}. \tag{3.10}$$

The solution of the system of equations (3.7) gives: $A_1 = 0$ and $A_3 = 0$. The coefficient $A_2 = A$ is found by the normalization of the function, $\psi_2(x) = A_2 \cosh[k\,(x - x_0)]\,e^{i\,G(x-x_0)}$, namely,

$\int_{x_0-d/2}^{x_0+d/2} A \cosh[k\,(x - x_0)]\,e^{i\,G(x-x_0)}\,A^* \cosh[k\,(x - x_0)]\,e^{-i\,G(x-x_0)}\,dx = \frac{A\,A^*[d\,K+\sin(d\,K)]}{2\,K} =$
$\frac{A\,A^*\,d}{2} = 1 \;\rightarrow\; |A| = \sqrt{2/d} = \sqrt{2\,K/q}$.

So, the normalized wave functions in the bound state:

$$\mathbf{\psi(x) = A \cos[\,K\,(x - x_0)]\,e^{i\,G(x-x_0)}}\; \text{or}\; \mathbf{\psi(x) = A \cos[\,K\,x]\,e^{i\,G(x)}} \tag{3.11}$$

**(c)** For function $\sinh[k\,(x - x_0)]$

According to the figure 1a, in the domain I, $E < U(x)$, in the domain II, $E > U(x)$, in the domain III, $E < U(x)$. According to the equations (2.20), the corresponding wave functions are written as follows:

$\psi_1(x) = A_1 \sinh[i\,k\,(x - x_0)]\,e^{i\,i\,G(x-x_0)}$ ; $\psi_2(x) = A_2 \sinh[k\,(x - x_0)]\,e^{i\,G(x-x_0)}$
$\psi_3(x) = A_3 \sinh[-i\,k\,(x - x_0)]\,e^{i\,i\,G(x-x_0)}$ ; $x_1 = x_0 - d/2$ and $x_2 = x_0 + d/2$

According to the conditions (3.1) we have:

$\psi_1(x_1) = \psi_2(x_1)$ or $\psi_1(x_1) - \psi_2(x_1) = 0$ ; $\psi_2(x_2) = \psi_3(x_2)$ or $\psi_2(x_2) - \psi_3(x_2) = 0$ (3.12)

From (3.12) it is obtained two linear equations as follows:

$$a_{11}A_1 + a_{12}A_3 = 0 \; ; \; a_{21}A_1 + a_{22}A_3 = 0 \tag{3.13}$$

Here, $a_{11} = -i\, e^{-G(-d/2)} \sin(d\,k/2)$ ; $a_{12} = 0$; $a_{21} = 0$ ; $a_{22} = i\, e^{G(d/2)} \sin(d\,k/2)$

In order this system of equations to have a solution different from zero, the determinant of coefficients should vanish, namely,

$$\det \begin{vmatrix} a_{11} & a_{12} \\ a_{21} & a_{22} \end{vmatrix} = 0 \quad \rightarrow \quad e^{-G(-d/2)+G(d/2)} \sin^2(d\,k/2) = 0 \tag{3.14}$$

In equation (3.14), $e^{-G(-d/2)+G(d/2)} \neq 0$, $k = m_1\sqrt{-E} = i\,m_1\sqrt{E} = i\,K$, $(E > 0, \; K = m_1\sqrt{E})$

So, $\sin^2(d\,k/2) = \sin^2(i\,d\,K/2) = \sin^2(d\,K/2) = 0$. From this last equation, we have that:

$$\mathbf{d\,K = q = 2\,n\,\pi, \; (n = 1,2,3,\ldots \text{ integer}).} \tag{3.15}$$

The solution of the system of equations (3.13) gives, $A_1 = 0$ and $A_3 = 0$. The coefficient $A_2 = A$ is found by the normalization of the function, $\psi_2(x) = A_2 \sinh[k\,(x-x_0)]\, e^{i\,G(x-x_0)}$, namely,

$$\int_{x_0-d/2}^{x_0+d/2} A \sinh[k\,(x-x_0)]\, e^{i\,G(x-x_0)} \, A^* \sinh[k\,(x-x_0)]\, e^{-i\,G(x-x_0)} \, dx = \frac{A\,A^*[-d\,K+\sin(d\,K)]}{2\,K} =$$

$$-\frac{A\,A^*\,d}{2} = 1 \quad \rightarrow \quad |A| = \sqrt{2/d} = \sqrt{2\,K/q}$$

So, the normalized wave functions in the bound state:

$$\psi(x) = A \sin[K\,(x-x_0)]\, e^{i\,G(x-x_0)} \text{ or } \psi(x) = A \sin[K\,x]\, e^{i\,G(x)} \tag{3.16}$$

The equations (3.10) and (3.15) can be combined as follows:

$$\mathbf{d\,K = q = n\,\pi, \; (n = 1,2,3,4,\ldots \text{ integer})} \tag{3.17}$$

So, in the bound states, the normalized wave functions are:

$$\psi(x) = \sqrt{2/d}\, \cos[K\,(x-x_0)]\, e^{i\,G(x-x_0)} \text{ and } \psi(x) = \sqrt{2/d}\, \sin[K\,(x-x_0)]\, e^{i\,G(x-x_0)} \tag{3.18a}$$

Or, $\psi(x) = \sqrt{2/d}\, \cos[K\,x]\, e^{i\,G(x)}$ and $\psi(x) = \sqrt{2/d}\, \sin[K\,x]\, e^{i\,G(x)}$ (3.18b)

The energy: $E_q = \frac{\hbar^2}{2\,m} \frac{q^2}{d^2} = M_h \frac{q^2}{d^2}, \; (M_h = \frac{\hbar^2}{2\,m})$ (3.18c)

Now, in bound states, let us see the relations between potential areas: (see figure 2)

**(a)** According to the partial integration; $\int u\, dv = u\,v - \int v\, du$, it can be written as follows:

$$S = S_p = \int_{x_1}^{x_2} U(x)dx = [x\, U(x)]_{x_1}^{x_2} - \int_{x_1}^{x_2} x\, U'(x)\, dx = x_2\, U(x_2) - x_1\, U(x_1) - S_t$$

$S_t = \int_{x_1}^{x_2} x\, U'(x)dx$ ; $U(x_1) = U(x_2) = E_q$ ; $d = x_2 - x_1$; Thus, it can be written:

$$S_p = E\,d - S_t \quad \text{or} \quad E\,d = S_p + S_t \tag{3.19a}$$

**(b)** Total energy = kinetic energy + potential energy; $E = T + U(x)$. From here, with integration:

$$\int_{x_1}^{x_2} E\, dx = \int_{x_1}^{x_2} T\, dx + \int_{x_1}^{x_2} U(x)\, dx \quad \rightarrow \quad E\,d = S_k + S_p \tag{3.19b}$$

If (3.19a) and (3.19b) are compared, it is seen that $S_k = S_t = \int_{x_1}^{x_2} x\, U'(x)dx$

**(c)** According to the equation (2.5): $E = -\frac{m}{2\,\hbar^2} S^2 = -\frac{m}{2\,\hbar^2} S_p^2 \quad \rightarrow \quad S_p = \sqrt{\frac{2\,\hbar^2}{m}} \, |E|$ ;

$$K\,d = q \;\to\; \sqrt{\tfrac{2m}{\hbar^2}|E|}\,d = q \;\to\; |E| = \tfrac{\hbar^2}{2m}\tfrac{q^2}{d^2} = M_h \tfrac{q^2}{d^2} \;\to\; S_p = 2\,M_h\,\tfrac{q}{d} \tag{3.19c}$$

**(d)** From (3.19a) and (3.19b) we have obtained: $\quad S_k = M_h \tfrac{q}{d}(q-2)$ (3.19d)

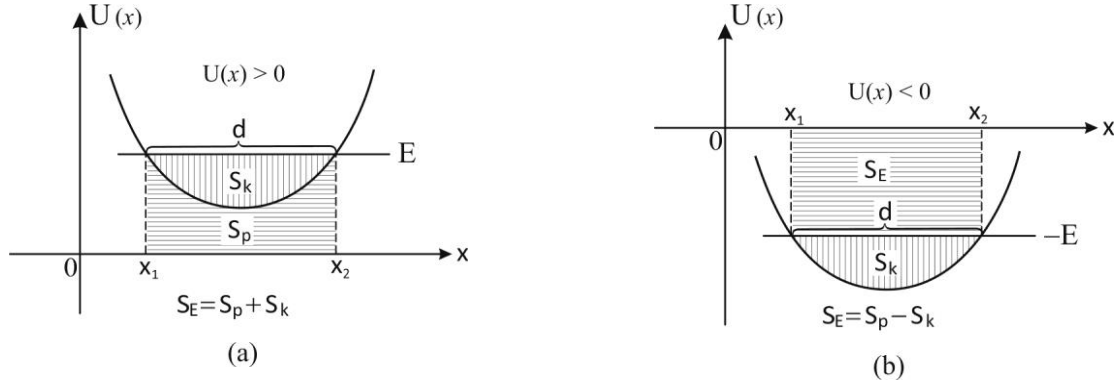

**Figure 2.** Potential area in the bound state: **(a)** $U(x) > 0$ and $E > 0$ ; **(b)** $U(x) < 0$ and $E < 0$

Let us rewrite these potential areas as follows:

$$S_p = \int_{x_1}^{x_2} U(x)\,dx = 2\,M_h\,\tfrac{q}{d} \quad \text{(Potential energy field)} \tag{3.20a}$$

$$S_k = \int_{x_1}^{x_2} x\,U'(x)\,dx = M_h\,\tfrac{q}{d}(q-2) \quad \text{(Kinetic energy field)} \tag{3.20b}$$

$$S_E = S_p + S_k = M_h\,\tfrac{q^2}{d} \quad \text{(Total energy field)} \tag{3.20c}$$

For $q = 2$, $S_k = 0$, the ground state occurs; for $q = n\pi$, ($n = 1, 2, 3, \dots$), the excited states occur. **The (3.20) equations indicate that potential areas are quantized.**

## 4. Summary

It is possible to summarize the above results briefly as follows.

We can write the general solutions of SE in one dimension as follows:

$$\psi(x) = A\,e^{kx}\,e^{\pm i\,G(x)} \quad \text{or} \quad \psi(x) = A\,e^{k(x-x_0)}\,e^{\pm i\,G((x-x_0))} \tag{4.1a}$$

$$\psi(x) = A\,e^{-kx}\,e^{\pm i\,G(x)} \quad \text{or} \quad \psi(x) = A\,e^{-k(x-x_0)}\,e^{\pm i\,G((x-x_0))} \tag{4.1b}$$

$$\psi(x) = \left[A\,e^{kx} + B\,e^{-kx}\right]e^{\pm i\,G(x)} \quad \text{or} \quad \psi(x) = \left[A\,e^{k(x-x_0)} + B\,e^{-k(x-x_0)}\right]e^{\pm i\,G(x-x_0)} \tag{4.1c}$$

$$\psi(x) = A\,\cosh(k\,x)\,e^{\pm i\,G(x)} \quad \text{or} \quad \psi(x) = A\,\cosh[k(x-x_0)]\,e^{\pm i\,G(x-x_0)} \tag{4.1d}$$

$$\psi(x) = A\,\sinh(k\,x)\,e^{\pm i\,G(x)} \quad \text{or} \quad \psi(x) = A\,\sinh[k(x-x_0)]\,e^{\pm i\,G(x-x_0)} \tag{4.1e}$$

$$\psi(x) = A\,e^{kx \pm i\,G(x)} + B\,e^{-kx \mp i\,G(x)} \quad \text{or} \quad \psi(x) = A\,e^{k(x-x_0)\pm i\,G(x)} + B\,e^{-k(x-x_0)\mp i\,G(x)} \tag{4.1f}$$

In these functions, we have the following values:

(a) For $E > U(x)$; $\quad k = m_1\sqrt{-E}$ , $\quad G(x) = m_1 \int \sqrt{-U(x)}\,dx$ (4.2a)

(b) For $E < U(x)$;  $k = m_1\sqrt{E}$ , $G(x) = m_1 \int \sqrt{U(x)}\, dx$ (4.2b)

Here, $m_1 = \sqrt{2m}/\hbar = \sqrt{2m/\hbar^2}$ , [m mass or reduced mass of particle, $\hbar = h/(2\pi)$, h Planck constant] ; $x_1$ and $x_2$ ($x_2 > x_1$) are, depending on E, the roots of the equation $E = U(x)$; and $x_0 = (x_1 + x_2)/2$ ; $d = x_2 - x_1$;

This solution is similar to the WKB approach but is not exactly the same. There is approximation in the WKB method, but there is no approach in the method we have given here. This procedure gives exact results. Those who are familiar with the WKB approach can easily see the differences.

In bound states, the normalized wave functions are as follows [G is taken as real function):

$\psi(x) = A \cos[Kx]\, e^{iG(x)}$ or $\psi(x) = A \cos[K(x - x_0)]\, e^{iG(x-x_0)}$ (4.3a)

$\psi(x) = B \sin[Kx]\, e^{iG(x)}$ or $\psi(x) = B \sin[K(x - x_0)]\, e^{iG(x-x_0)}$ (4.3b)

$A = B = \sqrt{2/d} = \sqrt{2K/q}$ ; $K = m_1\sqrt{|E|} = \sqrt{\frac{2m}{\hbar^2}}\sqrt{|E|}$ ; $G(x) = m_1 \int \sqrt{|U(x)|}\, dx$

$x_1$ and $x_2$ are the roots of the equation: $|E| = |U(x)|$ or $|E| = |U(x - x_0)|$. From this we have:

$x_0 = (x_1 + x_2)/2$ ; $d = x_2 - x_1$ ; $x_1 = x_0 - d/2$ ; $x_2 = x_0 + d/2$; $E = U(x)$ ;

$|E| = |U(x_1)| = |U(x_2)|$ or $|E|=|U(-d/2)|=|U(d/2)|$ ; $2|E|=|U(-d/2)|+|U(d/2)|$. (4.3c)

$Kd = q \rightarrow \sqrt{\frac{2m}{\hbar^2}}|E|\, d = q \rightarrow |E| = \frac{\hbar^2 q^2}{2m d^2} = M_h \frac{q^2}{d^2}$ ; $[M_h = \frac{\hbar^2}{2m}]$ (4.4)

We have, $q = 2$ for the ground state; $q = n\pi$, $(n = 1, 2, 3, ...)$ for the excited states.

The energy values are also given by the following formulas:

$E = -\frac{m}{2\hbar^2} S^2 = -\frac{m}{2\hbar^2} S_p^2$ ; $S_p = \sqrt{\frac{2\hbar^2}{m}|E|}$ ; $S = S_p(x_0, d) = 2 M_h \frac{q}{d}$ ; $S_k(x_0, d) = M_h \frac{q}{d}(q - 2)$

**Practical procedure to find the energy values**: As seen in the equation (4.4), the total energy values depend on d, so d value should be calculated. The d value can be calculated by one of the equations (4.3c), but to find the values of d and energy E, the practical procedure is as follows:

First solving the equation $|U(x)| = M_h q^2/y$ , $(y = d^2)$, it is found the $x_1$, $x_2$, $d = x_2 - x_1$, $d_2(y) = d * d$ values. After the equation, $y = d_2(y)$ is solved and y value is found. So the energy value is obtained as: $|E| = M_h \frac{q^2}{y} = M_h \frac{q^2}{d^2} = \frac{\hbar^2 q^2}{2m d^2}$ (4.5)

For $q = 2$ , the ground state occurs; for $q = n\pi$, $(n = 1, 2, 3, ...)$ the excited states occur.

## 5. Examples of one dimensional potentials

*5.1. Let us consider the potential*: $U(x) = a|x|^p$ , $(a > 0, p > 0)$

(a) **Energy:** According to the above practical procedure:

$E = U(x) = a|x|^p = M_h \frac{q^2}{y} \rightarrow x_1 = -\left[\frac{M_h}{a}\frac{q^2}{y}\right]^{1/p}$ ; $x_2 = \left[\frac{M_h}{a}\frac{q^2}{y}\right]^{1/p}$

$d = x_2 - x_1 = 2\left[\frac{M_h}{a}\frac{q^2}{y}\right]^{1/p}$ ; $d_2(y) = d * d = 4\left[\frac{M_h}{a}\frac{q^2}{y}\right]^{2/p}$ ; $y = d_2(y)$

$$y = \left[\frac{a\ 2^{-p}}{M_h\ q^2}\right]^{-2/(p+2)} \quad ; \quad E_q = M_h \frac{q^2}{y} = M_h\ q^2 \left[\frac{a\ 2^{-p}}{M_h\ q^2}\right]^{2/(p+2)} \quad (1)$$

For $q = 2$, the ground state occurs; for $q = n\pi$, $(n = 1, 2, 3, ...)$, the excited states occur.

In this formula (1), if it is taken $p = 2$ and $a = \frac{1}{2} m \omega^2$, it is found that $E_q = \frac{q}{4} \hbar \omega$. This is the energy of the simple harmonic oscillator. We have, $E_0 = \frac{1}{2} \hbar \omega$ ground state energy for $q = 2$; $E_q = \frac{n\pi}{2} \hbar \omega$, $(n = 1, 2, 3, ..)$ excited state energy for $q = n\pi$. The well-known energy of the simple harmonic oscillator is $E_n = \left(n + \frac{1}{2}\right) \hbar \omega$, $(n = 0, 1, 2, ...)$. Ground state energy $E_0 = \frac{1}{2} \hbar \omega$ is the same but excited energy $E_q = E_n = \frac{n\pi}{2} \hbar \omega$ is different [8]. I believe that this latter is more accurate because there is no approximation here.

For the same potential, the ground state energy obtained from Supersymmetric Quantum Mechanics (SQ) is as follows [10]: 
$$E_0 = \left[\frac{0.8862\ \Gamma\left(\frac{3}{2}+\frac{1}{p}\right)}{\Gamma\left(1+\frac{1}{p}\right)} \frac{\hbar^2}{2m} a^{\frac{2}{p}}\right]^{p/(p+2)} \quad (2)$$

From this equation (2), for $p = 2$ and $a = \frac{1}{2} m \omega^2$, we find that $E_0 = 0.999985 \frac{1}{2} \hbar \omega \approx \frac{1}{2} \hbar \omega$.

In this potential, for $p = 1$, (V-Form potential), it is obtained $E_0 \approx 0.794 \left[\frac{\hbar^2 a^2}{m}\right]^{1/3}$ from the formula (1); $E_0 \approx 0.763 \left[\frac{\hbar^2 a^2}{m}\right]^{1/3}$ from SQ [10]; $E_0 \approx 0.813 \left[\frac{\hbar^2 a^2}{m}\right]^{1/3}$ from variation method [11]; and $E_0 \approx 0.885 \left[\frac{\hbar^2 a^2}{m}\right]^{1/3}$ from WKB method [11]. The three values are approximately the same.

**(b) Wave Functions:**

$$G(x) = m_1 \int \sqrt{|U(x)|}\ dx = \sqrt{\frac{2m}{\hbar^2}} \int \sqrt{U(x)}\ dx = Q(x) = \sqrt{\frac{2m}{\hbar^2}} \sqrt{a}\ \frac{2\ |x|^{(p+2)/2}}{p+2}$$

$$\psi(x) = A \cos[K x]\ e^{i\ G(x)} = \sqrt{2 K/q}\ \cos[K x]\ e^{i\ Q(x)} = \sqrt{2/d}\ \cos[K x]\ e^{i\ Q(x)} \quad (3a)$$

$$\psi(x) = B \sin[K x]\ e^{i\ G(x)} = \sqrt{2 K/q}\ \sin[K x]\ e^{i\ Q(x)} = \sqrt{2/d}\ \sin[K x]\ e^{i\ Q(x)} \quad (3b)$$

In the case of simple harmonic oscillator: $Q(x) = \sqrt{\frac{2m}{\hbar^2}} \sqrt{a}\ \frac{2\ |x|^{(p+2)/2}}{p+2} = \frac{m\omega}{2\hbar} x^2$. The well-known wave function of the harmonic oscillator is: $\psi_n(\rho) = A_n e^{-\rho^2/2} H_n(\rho)$, $[\rho = \sqrt{\frac{m\omega}{\hbar}} x]$ (4)

Here, $H_n(\rho)$ is Hermit Polynomials. If the equations (3a) and (3b) are expanded in series, polynomials are obtained. So the well-known function (4) is an approximate function.

*5.2. Infinitely high square potential well or finite square potential well*

We consider a particle of mass m captured in a box limited by, $0 \le x \le a$. The corresponding potential is given: $U(x) = 0$ for $0 < x < a$; $U(x) = \infty$ for $x < 0$ and $x > a$.

**(a) Energy:** The turning points of this potential are given by the following equation: $U(x) = E$; From this equation, we can find the classical turning points of the potential function as:

$x_1 = 0$ and $x_2 = a$; $x_0 = (x_1 + x_2)/2 = a/2$, $d = x_2 - x_1 = a$

By substituting this value of d into the equation, $dK = q$, and then solving for $|E|$, we can have the energy value as: $E_q = \frac{\hbar^2}{2m} \frac{q^2}{a^2} = M_h \frac{q^2}{a^2}$ ; for $q = 2$ the ground state; for $q = n\pi$, $(n = 1, 2, 3, ...)$ the excited states. The well-known energy is $E_n = \frac{\hbar^2}{2m} \frac{(n\pi)^2}{a^2}$ [8]. Two energy values are the same.

**(b) Wave Functions:** Here, $U(x) = 0$, so $G(x) = m_1 \int \sqrt{|U(x)|}\, dx = 0$

$\psi(x) = A \cos[Kx]\, e^{i\, G(x)} = \sqrt{2/a}\, \cos[Kx] = \sqrt{2K/q}\, \cos[Kx]$

$\psi(x) = B \sin[Kx]\, e^{i\, G(x)} = \sqrt{2/a}\, \sin[Kx] = \sqrt{2K/q}\, \sin[Kx]$

$\psi(x) = A \cos\left[K(x - \frac{a}{2})\right] = \sqrt{2/a}\, \cos\left[K(x - \frac{a}{2})\right] = \sqrt{2K/q}\, \cos\left[K(x - \frac{a}{2})\right]$

$\psi(x) = B \sin\left[K(x - \frac{a}{2})\right] = \sqrt{2/a}\, \sin\left[K(x - \frac{a}{2})\right] = \sqrt{2K/q}\, \sin\left[K(x - \frac{a}{2})\right]$

*5.3. Trigonometric potential well*

We consider the potential energy of the particle, $U(x)$, so that $U(x) = U_0\, \cotg^2(\pi x/a)$, $[U_0 > 0,\ a > 0$ and $0 < x < a]$.

**(a) Energy:** The roots of the equation $E_q = U(x) = M_h \frac{q^2}{y} \rightarrow U_0 \cotg^2[\pi x/a] = M_h \frac{q^2}{y}$

$x_1 = -\frac{1}{\pi} \text{arc cotg}\left[\sqrt{\frac{M_h\, q^2}{U_0\, y}}\right]$ ; $x_2 = \frac{1}{\pi} \text{arc cotg}\left[\sqrt{\frac{M_h\, q^2}{U_0\, y}}\right]$ ; $d = x_2 - x_1$ ; $y = d * d$

$y = \frac{4}{\pi^2} \text{arc cotg}^2\left[\sqrt{\frac{M_h\, q^2}{U_0\, y}}\right]$ ; $E_q = M_h \frac{q^2}{d^2} = M_h \frac{q^2}{y}$ (1)

The equation of (1) is not solved analytically, so it may be solved by the numerical method for the exact values. We have, for $q = 2$ the ground state; for $q = n\pi$, $(n = 1, 2, 3, ...)$ the excited states.

**(b) Wave Functions:**

$G(x) = m_1 \int \sqrt{|U(x)|}\, dx = m_1 \int \sqrt{U(x)}\, dx = m_1 \int \sqrt{U_0\, \cotg^2(\pi x/a)}\, dx = Q(x)$

$Q(x) = \sqrt{\frac{2m a^2 U_0}{\hbar^2 \pi^2}}\, \ln\left[\sin(\frac{\pi x}{a})\right]$ (2)

$\psi(x) = A \cos[Kx]\, e^{i\, G(x)} = \sqrt{2K/q}\, \cos[Kx]\, e^{i\, Q(x)} = \sqrt{2/d}\, \cos[Kx]\, e^{i\, Q(x)}$ (3a)

$\psi(x) = B \sin[Kx]\, e^{i\, G(x)} = \sqrt{2K/q}\, \sin[Kx]\, e^{i\, Q(x)} = \sqrt{2/d}\, \sin[Kx]\, e^{i\, Q(x)}$ (3b)

*5.4. Infinitely high parabolic potential well*

We consider the potential energy of the particle, $U(x)$, so that $U(x) = U_0 \left(\frac{a}{x} - \frac{x}{a}\right)^2$, $[U_0 > 0,\ a > 0,\ x > 0]$.

**(a) Energy:** The positive roots of the equation of $E = U(x) = U_0 \left(\frac{a}{x} - \frac{x}{a}\right)^2 = M_h \frac{q^2}{y}$ :

$x_1 = \sqrt{\frac{a^2\left[M_h q^2 + 2 y U_0 - q \sqrt{M_h}\sqrt{M_h q^2 + 4 y U_0}\right]}{2 y U_0}}$ ; $x_2 = \sqrt{\frac{a^2\left[M_h q^2 + 2 y U_0 + q \sqrt{M_h}\sqrt{M_h q^2 + 4 y U_0}\right]}{2 y U_0}}$

$$d_2 = d^2 = (x_2 - x_1)^2 =$$

$$\frac{a}{2} \left[ \sqrt{\frac{[M_h q^2 + 2 y U_0 + q \sqrt{M_h} \sqrt{M_h q^2 + 4 y U_0}]}{y U_0}} - \sqrt{\frac{[M_h q^2 + 2 y U_0 - q \sqrt{M_h} \sqrt{M_h q^2 + 4 y U_0}]}{y U_0}} \right]^2$$

The root of the equation of $d_2 = y$ is $y = \sqrt{a^2 \frac{M_h q^2}{U_0}}$ and $E_q = M_h \frac{q^2}{d^2} = M_h \frac{q^2}{y} = \sqrt{\frac{M_h U_0}{a^2}} \, q$

$$E_q = \sqrt{\frac{\hbar^2 U_0}{2 m a^2}} \, q \tag{1}$$

We have the ground state for $q = 2$; the excited states for $q = n\pi$, $(n = 1, 2, 3, ...)$ [8].

For $m = 1$ and $\hbar = 1$, this energy value is: $E_n = \frac{\pi}{2} n \sqrt{\frac{2 U_0}{a^2}}$

For this potential, the energy values obtained from Supersymmetric WKB [12] and Standard WKB [14], respectively, are as follows:

$$E_n = \sqrt{\frac{2 U_0}{a^2}} \left[ 2 n + \sqrt{1 + 8 a^2 U_0} \right] - 2 U_0 \quad \text{and} \quad E_n = \sqrt{\frac{2 U_0}{a^2}} \left[ 2 n + 1 + \sqrt{2 a^2 U_0} \right] - 2 U_0$$

**(b) Wave Functions:**

$$G(x) = m_1 \int \sqrt{|U(x)|} \, dx = m_1 \int \sqrt{U(x)} \, dx = m_1 \int \sqrt{U_0 \left(\frac{a}{x} - \frac{x}{a}\right)^2} \, dx = m_1 \int \sqrt{U_0} \left(\frac{a}{x} - \frac{x}{a}\right) dx =$$

$$Q(x) ; \quad Q(x) = \sqrt{\frac{2 m U_0}{\hbar^2}} \left[ a \ln(x) - \frac{x^2}{2 a} \right]$$

$$\psi(x) = A \cos[K x] \, e^{\pm i Q(x)} ; \quad \psi(x) = B \sin[K x] \, e^{\pm i Q(x)} ; \quad |A| = |B| = \sqrt{2/d} = \sqrt{2 K/q} \tag{2}$$

## 5. 5. Potential $U(x) = a x^2 + b/x^2$

We consider the potential energy of the particle as $U(x) = a x^2 + b/x^2$, here a and b are positive constants.

**(a) Energy:** The roots of the equation of $U(x) = a x^2 + b/x^2 = M_h \frac{q^2}{y} = E_q$ are as follows:

$$x_1 = -\sqrt{\frac{E_q + \delta}{2 a}}, \quad x_2 = \sqrt{\frac{E_q - \delta}{2 a}}, \quad x_3 = -\sqrt{\frac{E_q - \delta}{2 a}}, \quad x_4 = \sqrt{\frac{E_q + \delta}{2 a}} \; ; \; [\delta = \sqrt{E_q^2 - 4 a b} \,].$$

From these grandeurs, we get,

$$d_{31} = x_3 - x_1 = -\sqrt{\frac{E_q + \delta}{2 a}} + \sqrt{\frac{E_q - \delta}{2 a}} \; ; \; d_{31}^2 = E_q/a - 2 \sqrt{b/a}$$

$$d_{42} = x_4 - x_2 = \sqrt{\frac{E_q + \delta}{2 a}} - \sqrt{\frac{E_q - \delta}{2 a}} \; ; \; d_{42}^2 = E_q/a - 2 \sqrt{b/a}$$

From these equations, we can easily obtain: $d_{31} = -d_{42} = d$, $d^2 = E_q/a - 2\sqrt{b/a}$

From the solution of the equation of $E_q = M_h \frac{q^2}{d^2}$, we get the positive energy value as follows:

$$E_q = \sqrt{a b} + \sqrt{a b + a M_q q^2} = \sqrt{a b} + \sqrt{a b + a \frac{\hbar^2}{2 m} q^2} \tag{1}$$

For $q = 2$ the ground state; for $q = n\pi$, $(n = 1, 2, 3, ...)$ the excited states.

For $m = 1$ and $\hbar = 1$, this energy becomes:

$$E_q = \sqrt{a\,b} + \sqrt{a\,b + \frac{a}{2}\pi^2\,n^2}\,, \quad (n = 1, 2, 3, \dots) \tag{2a}$$

The energies obtained from Supersymmetric WKB and Standard WKB [8, 12, 13, 14], respectively, are

$$E_n = \sqrt{2\,a}\,\left[2\,n + 1 + \sqrt{\frac{1}{4} + 2\,b}\right], \quad (n = 0, 1, 2, 3, \dots) \tag{2b}$$

$$E_n = \sqrt{2\,a}\,[2\,n + 1 + \sqrt{2\,b}], \quad (n = 0, 1, 2, 3, \dots) \tag{2c}$$

It can be assumed that (1) and (2a) values are more accurate because there is no approximation.

**(b) Wave Functions:**

$$G(x) = m_1 \int \sqrt{|U(x)|}\,dx = m_1 \int \sqrt{U(x)}\,dx = m_1 \int \sqrt{a\,x^2 + b/x^2}\,dx = m_1 \frac{1}{2}\{\sqrt{b + a\,x^4} + \sqrt{b}\,[\ln(x^2) - \ln(b + \sqrt{b}\,\sqrt{b + a\,x^4}\,)]\} = Q(x)$$

$$Q(x) = \sqrt{\frac{m}{2\,\hbar^2}}\,\{\sqrt{b + a\,x^4} + \sqrt{b}\,[\ln(x^2) - \ln(b + \sqrt{b}\,\sqrt{b + a\,x^4}\,)]\}$$

$$\psi(x) = A \cos[K\,x]\,e^{i\,Q(x)} \text{ and } \psi(x) = B \sin[K\,x]\,e^{i\,Q(x)}\,;\ |A| = |B| = \sqrt{2/d} = \sqrt{2\,K/q} \tag{3}$$

## 6. Radial Schrödinger equation for spherical symmetric potentials

Schrödinger Equation (SE) in three dimensions is given as follows,

$$\Delta\Psi(\vec{r}) + \frac{2\,m}{\hbar^2}\,[E - V(\vec{r})]\,\Psi(\vec{r}) = 0 \tag{6.1}$$

Where, E and V are the total and potential energies, respectively, m is the mass or reduced mass of particle. Spherical polar coordinates $x = r\sin(\theta)\cos(\phi)$, $y = r\sin(\theta)\sin(\phi)$, $z = r\cos(\theta)$ are appropriate for the symmetry of the problem. The SE (6.1), expressed in these coordinates, is

$$\left[\frac{\partial^2}{\partial r^2} + \frac{2}{r}\frac{\partial}{\partial r}\right]\Psi(r,\theta,\phi) + \frac{1}{r^2}\hat{\mathcal{L}}^2(\theta,\phi)\Psi(r,\theta,\phi) + \frac{2\,m}{\hbar^2}\,[E - V(r,\theta,\phi)]\Psi(r,\theta,\phi) = 0 \tag{6.2a}$$

Here, $\hat{\mathcal{L}}^2(\theta,\phi) = \frac{\partial^2}{\partial\theta^2} + \cotg(\theta)\frac{\partial}{\partial\theta} + \frac{1}{\sin^2(\theta)}\frac{\partial^2}{\partial\phi^2}$ \hfill (6.2b)

The potential energy of a particle which moves in a central, spherically symmetric field of force depends only upon the distance r between the particle and the force center. Thus, the potential energy should be $V(r,\theta,\phi) = V(r)$. Solution of the Eq. (6.2) can be found by the method of separation of variables. To apply this method, it is assumed that the solution is in the following form:

$$\Psi(r,\theta,\phi) = R(r)\,Y(\theta,\phi) \text{ or } \Psi(r,\theta,\phi) = R(r)\,|j\,m> \tag{6.3}$$

In (6.3), $R(r)$ is independent of the angles and $Y(\theta,\phi)$ or $|jm>$ is independent of r. Substituting Eq. (6.3) into equation (6.2a) and rearranging, the following two equations are obtained:

$$\frac{\partial^2 R(r)}{\partial r^2} + \frac{2}{r}\frac{\partial R(r)}{\partial r} + \left\{\frac{2\,m}{\hbar^2}\,[E - V(r)] - \frac{C}{r^2}\right\}R(r) = 0 \tag{6.4}$$

$$\hat{\mathcal{L}}^2(\theta,\phi)\,Y(\theta,\phi) + C\,Y(\theta,\phi) = 0 \tag{6.5}$$

Where C is constant. Eq. (6.5) is independent of the total energy E and of the potential energy $V(r)$, therefore, the angular dependence of the wave functions is determined by the property of spherical symmetry and admissible solutions of Eq. (6.5) are valid for every spherically symmetric system regardless of the special form of the potential function. The solutions of the Eq. (6.5) can be found in any quantum mechanics and mathematical physics text–books [15-18] and the solutions are known as spherical harmonic functions, $Y_{\ell\mu}(\theta,\phi)$, where $C = \ell(\ell + 1), (\ell = 0, 1, 2, 3, \dots)$ are positive integer numbers and $\mu = -\ell, -\ell+1, -\ell+2, \ldots 0, 1, 2, \ldots, \ell$. Eq. (6.4) is the radial SE. Substituting $C = \ell(\ell+1)$ and $F(r) = r\,R(r)$ values into Equation (6.4), the radial wave equation is obtained as:

$$\frac{\partial^2 F(r)}{\partial r^2} + \frac{2m}{\hbar^2}[E - U(r)]F(r) = 0 \tag{6.6}$$

Here, $U(r) = V(r) + \frac{\hbar^2}{2m}\frac{\ell(\ell+1)}{r^2}$ is the effective potential energy. This equation (6.6) is one dimensional differential equation and is the same as (1.2) equation. In (1.2) free variable x, in (6.6) free variable is r. So the solution procedure of one dimensional differential equation has been given above (4.1-4.5). Now let us give some examples.

### 6.1. Coulomb type central potential well

The potential energy of hydrogen-like atom is: $V(r) = -Z\,e^2/r$. If it is added the centrifugal potential function to this potential $V(r)$, we have the following effective potential function:

$$U(r) = -\frac{a}{r} + \frac{b}{r^2}, \quad \left[a = Z\,e^2 \text{ and } b = \frac{\hbar^2}{2m}\frac{\ell(\ell+1)}{r^2}\right].$$

**(a) Energy:** The classical turning points of this effective potential are given by the following equation, $-\frac{a}{r} + \frac{b}{r^2} = -|E| = -M_h \frac{q^2}{y}$. From this equation, we can find the classical turning points of this effective potential function and $d_2$ as [21]:

$$r_1 = \frac{a\,y - \sqrt{a^2 y^2 - 4\,b\,M_h\,y\,q^2}}{2\,M_h q^2} \; ; \; r_2 = \frac{a\,y + \sqrt{a^2 y^2 - 4\,b\,M_h\,y\,q^2}}{2\,M_h q^2} \text{ and } d_2 = (r_2 - r_1)^2 = \frac{y\,[y\,a^2 - 4\,b\,M_h q^2]}{M_h^2\,q^4}$$

The root of the equation of $d_2 = y$ is $y = \frac{M_h q^2\,[M_h q^2 + 4\,b]}{a^2}$ and $|E_q| = M_h \frac{q^2}{d^2} = M_h \frac{q^2}{y}$; thus

$$E_q = -\frac{2\,m\,e^4\,Z^2}{\hbar^2\,[4\,\ell + 4\,\ell^2 + q^2]} = -\frac{m\,e^4}{2\,\hbar^2}\frac{Z^2}{[\ell(\ell+1) + q^2/4]} = -E_0\frac{Z^2}{[\ell(\ell+1) + q^2/4]}, \quad [E_0 = \frac{m\,e^4}{2\,\hbar^2}] \tag{1}$$

We have the ground state for $q = 2$; the excited states for $q = n\,\pi$, $(n = 1, 2, 3, \ldots)$.

For the hydrogen atom $Z = 1$ and m electron mass, e electron charge, in the ground state, $\ell = 0$. From the equation (1), $E_0 = -13.6$ eV is obtained. This is the well-known ground state energy of the hydrogen atom. The well-known excited state energy of hydrogen-like atoms is given as follows:

$E_n = -E_0\,\frac{Z^2}{n^2}$, $(n = 1, 2, 3, \ldots)$. There is no obviously $\ell = 0$ quantum number in this formula, but in the formula (1) this number of quantum is clearly visible [19].

**(b) Wave Functions:**

$$G(r) = m_1 \int \sqrt{|U(r)|}\,dr = \sqrt{\frac{2m}{\hbar^2}}\int \sqrt{-U(r)}\,dr = \sqrt{\frac{2m}{\hbar^2}}\int \sqrt{\frac{a}{r} - \frac{b}{r^2}}\,dr = Q(r);$$

$$Q(r) = 2\sqrt{\frac{2m}{\hbar^2}}\left[\sqrt{a\,r - b} - \sqrt{b}\,\arctan\!\left(\sqrt{\frac{a\,r-b}{b}}\right)\right]$$

$$F(r) = A\cos[K\,r]\,e^{i\,Q(r)} \text{ and } F(r) = \sqrt{\frac{2}{d}}\cos[K\,r]\,e^{i\,Q(r)} = \sqrt{\frac{2K}{q}}\cos[K\,r]\,e^{i\,Q(r)}$$

$$F(r) = B\sin\!\left[\frac{q}{d}r\right]e^{i\,Q(r)} \text{ and } F(r) = \sqrt{\frac{2}{d}}\sin\!\left[\frac{q}{d}r\right]e^{i\,Q(r)} = \sqrt{\frac{2K}{q}}\sin\!\left[\frac{q}{d}r\right]e^{i\,Q(r)}$$

$$\Psi(r, \theta, \phi) = R(r)\,Y(\theta, \phi) = \frac{F(r)}{r}\,Y(\theta, \phi) = \frac{F(r)}{r}\,|jm\rangle \tag{2}$$

In equation (2) m is not mass, it is magnetic quantum number. The known wave function of the hydrogen atom is an exponential function including Laguerre polynomials. Our results are more accurate because there is no approach.

**6.2. Infinitely high spherical symmetric square well or finite spherical symmetric square well**

Consider a particle of mass m captured in a box limited by $0 \leq r \leq a$. The corresponding central potential can be given by, $V(r) = 0$ for $0 \leq r \leq a$ ; $V(r) = \infty$ for $r < 0$ and $r > a$. With this potential, the effective potential: $U(r) = \frac{b}{r^2}$ ; $[b = \frac{\hbar^2}{2\,m}\ell(\ell+1)]$.

**(a) Energy:** With this effective potential, the equation $U(r) = \frac{b}{r^2} = \frac{M_h\, q^2}{y}$ can be written. From this equation, the classical turning points and some grandeur are found as follows [19]:

$$r_1 = \frac{1}{q}\sqrt{\frac{b\,y}{M_h}} \; ; \; r_2 = a, \; ; \; d_2 = (r_2 - r_1)^2 = \left[a - \frac{1}{q}\sqrt{\frac{b\,y}{M_h}}\right]^2 \qquad (1)$$

From the solution of the equation of $d_2 = y$; we can obtain $y = \frac{M_h\, a^2 q^2}{[\sqrt{b} \pm \sqrt{M_h}\, q]^2}$ and the energy value as:

$$E_q = \frac{[\sqrt{b} \pm \sqrt{M_h}\, q]^2}{a^2} = \frac{\hbar^2}{2\,m\,a^2}\left[\sqrt{\ell(\ell+1)} \pm q\right]^2 \qquad (2)$$

We have; for $q = 2$ the ground state; for $q = n\pi$, $(n = 1, 2, 3, \ldots)$, the excited states.

The known allowed energies are given as follows [9]: $E_{n\ell} = \frac{\hbar^2}{2\,m\,a^2}\beta_{n\ell}^2$ (3)

Here, $\beta_{n\ell}$, $n^{th}$ zero of the $\ell^{th}$ spherical Bessel functions. The equation (2) can be written as follows:

$$E_q = E_{n\ell} = \frac{\hbar^2}{2\,m\,a^2}\left[\sqrt{\ell(\ell+1)} \pm n\pi\right]^2 \qquad (4)$$

Some values of $\beta_{n\ell}$ and energies calculated according to (3) and (4) [with sign + in (4)] are given in the table 1.

**Table 1.** Some Energy Values of the Infinitely High Spherical Symmetric Square Well (Unit: $\hbar^2/(2\,m\,a^2)$).

| $n\ell$ | $\beta_{n\ell}$ | $E_{n\ell}$ From (3) | $E_{n\ell}$ From (4) |
|---|---|---|---|
| 1 s | 3.142 | 9.872 | 9.870 |
| 1 p | 4.493 | 20.187 | 20.755 |
| 1 d | 5.763 | 33.212 | 31.260 |
| 2 s | 6.283 | 39.476 | 39.478 |
| 2 p | 7.725 | 59.676 | 59.250 |
| 2 d | 9.095 | 82.719 | 76.260 |

**(b) Wave Functions:**

$$G(r) = m_1 \int \sqrt{|U(r)|}\, dr = \sqrt{\frac{2\,m}{\hbar^2}} \int \sqrt{|U(r)|}\, dr = m_1 \int \sqrt{\frac{b}{r^2}}\, dr = \sqrt{\frac{2\,m}{\hbar^2}} \int \sqrt{\frac{b}{r^2}}\, dr$$

$$G(r) = m_1 \int \sqrt{\frac{b}{r^2}}\, dr = \sqrt{\frac{2\,m}{\hbar^2}}\sqrt{b}\,\ln(r) = \sqrt{\ell(\ell+1)}\,\ln(r) = Q(r)$$

$$F(r) = = A\cos[K\,r]\, e^{i\,G(r)} = A\cos[K\,r]\, e^{i\,Q(r)} \;;\; F(r) = B\sin[K\,r]\, e^{i\,G(r)} = \sin[K r]]\, e^{i\,Q(r)}$$

$$F(r) = = A\cos\left[\frac{q}{d}\,r\right] e^{i\,G(r)} = A\cos\left[\frac{q}{d}\,r\right] e^{i\,Q(r)} \;;\; F(r) = B\sin\left[\frac{q}{d}\,r\right] e^{i\,G(r)} = \sin\left[\frac{q}{d}\,r\right] e^{i\,Q(r)}$$

$|A| = |B| = \sqrt{2/d} = \sqrt{2\,K/q}$ ;  $\Psi(r, \theta, \phi) = R(r)\,Y(\theta, \phi) = \frac{F(r)}{r}\,Y(\theta, \phi)$

## 6.3. Three-dimensional isotropic harmonic oscillator potential

The potential energy of three dimensional isotropic harmonic oscillators is given by:  $V(r) = \frac{1}{2}\,m\,\omega^2 r^2$.

With this potential; the effective potential $U(r)$ is as follows:

$U(r) = \frac{1}{2}\,m\,\omega^2 r^2 + \frac{\hbar^2 \ell(\ell+1)}{2\,m}\,\frac{1}{r^2} = a\,r^2 + \frac{b}{r^2}$ ; $[a = \frac{1}{2}\,m\,\omega^2$ and $b = \frac{\hbar^2 \ell(\ell+1)}{2\,m}]$

**(a) Energy:** From this effective potential, the positive roots of the equation $E_q = M_h \frac{q^2}{y} = U(r)$ are:

$$r_1 = \sqrt{\frac{E_q - \delta}{2\,a}} \;;\; r_2 = \sqrt{\frac{E_q + \delta}{2\,a}}; \quad [\,E_q = M_h \frac{q^2}{y}\,;\; \delta = \sqrt{E_q^2 - 4\,a\,b}\,] \tag{1}$$

With these roots; $d = r_2 - r_1 = \sqrt{\frac{E_q + \delta}{2\,a}} - \sqrt{\frac{E_q - \delta}{2\,a}}$ ;  $d_2 = y = d * d = d^2$ ; $\tag{2}$

By substituting this value of $d_2$ into the equation, $y = M_h \frac{q^2}{y}$ and then solving for $E_q$, we can obtain the appropriate energy as [19]:  $E_q = \frac{1}{2}\hbar\omega[\sqrt{\ell(\ell+1)} + \sqrt{\ell(\ell+1) + q^2}\,]$ $\tag{3}$

We have, for $q = 2$ the ground state; for $q = n\,\pi$, $(n = 1, 2, 3, ...)$ the excited states.

The known energy values are given as follows [20]:

$\quad E_{n\ell} = (2\,n + \ell + 3/2)\,\hbar\omega$ , $(n = 0, 1, 2, 3, ....)$ $\tag{4}$

Some values of the energies, calculated according to (3) and (4) are given in the table 2.

**Table 2.** Some energy values of the isotropic harmonic oscillator (unit $\hbar\omega$).

| n ℓ | $E_{n\ell}$ [according to (4)] | $E_{n\ell}$ [according to (3)] |
|---|---|---|
| 1 s | 3.500 | 1.571 |
| 1 p | 4.500 | 2.430 |
| 1 d | 5.500 | 3.217 |
| 2 s | 5.500 | 3.142 |
| 2 p | 6.500 | 3.927 |
| 2 d | 7.500 | 4.597 |

**(b) Wave Functions:**

$G(r) = m_1 \int \sqrt{|U(r)|}\,dr = \sqrt{\frac{2\,m}{\hbar^2}} \int \sqrt{U(r)}\,dr = \sqrt{\frac{2\,m}{\hbar^2}} \int \sqrt{a\,r^2 + \frac{b}{r^2}}\,dr = Q(r)$

$Q(r) = m_1 \frac{1}{2}\left\{\sqrt{a\,r^4 + b} - \sqrt{b}\,\ln\left[2\,\frac{\sqrt{b} + \sqrt{a\,r^4 + b}}{r^2}\right]\right\}$

$F(r) = A\cos[K\,r]\,e^{i\,G(r)} = A\cos[K\,r]\,e^{i\,Q(r)} = A\cos\left[\frac{q}{d}\,r\right]e^{i\,Q(r)}$ $\tag{5a}$

$F(r) = B\sin[K\,r]\,e^{i\,G(r)} = B\sin[K\,r]\,e^{i\,Q(r)} = B\sin\left[\frac{q}{d}\,r\right]e^{i\,Q(r)}$ $\tag{5b}$

$|A| = |B| = \sqrt{2/d} = \sqrt{2\,K/q}$ ;  $\Psi(r, \theta, \phi) = R(r)\,Y(\theta, \phi) = \frac{F(r)}{r}\,Y(\theta, \phi)$ $\tag{6}$

The known radial wave functions are [20]:  $R_{n\ell}(\rho) = A\,\rho^{\ell+1} e^{-\frac{1}{2}\rho^2}\,\mathcal{L}_n^{\ell+1/2}(\rho^2)$ $\tag{7}$

Where $\rho = \sqrt{m\,\omega/\hbar}\;r$ and $\mathcal{L}$ is the Laguerre polynomial.

*6.4. Three-dimensional isotropic harmonic oscillator with spin-orbit coupling. (Shell model in nuclear physics)*

The simple potential energy of a three dimensional isotropic harmonic oscillator is given by, $V_0(r) = \frac{1}{2} m \omega^2 r^2$. The spin–orbit interaction potential must be added to this simple potential. The spin–orbit interaction potential is given by: $V_{\ell sj}(r) = -\frac{\hbar^2}{2 m^2 c^2} \frac{1}{r} \frac{dV_0(r)}{dr} \vec{\ell}.\vec{s}$ ; $\{\vec{\ell}.\vec{s} = \frac{1}{2}[j(j+1) - \ell(\ell+1) - s(s+1)]\}$, here $\ell$, s and j are the orbital, spin and total angular momentum quantum numbers of a particle, respectively. It is possible to find the derivation of this expression in any quantum mechanics text book. If $V_{\ell sj}(r)$ is calculated, the following value is found:

$V_{\ell sj}(r) = -\frac{\hbar^2 \omega^2}{4 m c^2}[j(j+1) - \ell(\ell+1) - s(s+1)] = -C_{\ell sj}$. If s = 1/2 (for fermions) is accepted,

$V_{\ell sj}(r) = -\frac{\hbar^2 \omega^2}{4 m c^2}\left[j(j+1) - \ell(\ell+1) - \frac{3}{4}\right] = -C_{\ell sj}$ is obtained. Thus, the harmonic oscillator potential and effective potential with spin-orbit coupling becomes:

$V(r) = V_0(r) + V_{\ell sj}(r)$ ; $U(r) = a r^2 - C_{\ell sj} + b/r^2$ ; $\left[a = \frac{1}{2} m \omega^2, b = \frac{\hbar^2}{2 m} \ell(\ell+1)\right]$

**(a) Energy:** From this effective potential, the positive roots of the equation $E_q = M_h \frac{q^2}{y} = U(r)$ are:

$$r_1 = \sqrt{\frac{(E_q + C_{\ell sj}) - \delta}{2 a}} \; ; \; r_2 = \sqrt{\frac{(E_q + C_{\ell sj}) + \delta}{2 a}} \; ; \; \left[\delta = \sqrt{(E_q + C_{\ell sj})^2 - 4 a b}\right] \tag{1}$$

With these roots; $d = r_2 - r_1 = \sqrt{\frac{E_q + \delta}{2 a}} - \sqrt{\frac{E_q - \delta}{2 a}}$ ; $d_2 = y = d * d = d^2$ ; (2)

By substituting this value of $d_2$ into the equation, $y = M_h \frac{q^2}{y}$ and then solving for $E_q$, we can obtain the appropriate energy as follows:

$$E_q = \frac{1}{2} \hbar \omega \left[\sqrt{\ell(\ell+1)} - C_j + \sqrt{(\sqrt{\ell(\ell+1)} - C_j)^2 + q^2}\right], \; [C_j = C_{\ell sj}/(\hbar \omega)]. \tag{3}$$

We have; for q = 2 the ground state; for q = n π, (n = 1, 2, 3, ...) the excited states.

The known energy values (with first order perturbation) are as follows [7, 9, 20]:

$$E_{n\ell} = (2n + \ell + 3/2)\hbar \omega - \frac{C_0}{4 \hbar \omega}[j(j+1) - \ell(\ell+1) - s(s+1)]\hbar \omega \tag{4}$$

Here, n = 0, 1, 2,.., integer numbers and $C_0$ is a positive parameter. In table 3, some energy values calculated according to (3) and (4) are given, with the values:

$C_0 = 0.015 \hbar \omega$, s = 1/2 and $C_j = \frac{C_0}{4 \hbar \omega}[j(j+1) - \ell(\ell+1) - 3/4]\hbar \omega$

**Table 3.** Some energy values of the isotropic harmonic oscillator with spin-orbit coupling.

| States | According to (4) | (unit $\hbar\omega$) | According to (3) | (unit $\hbar\omega$) |
|---|---|---|---|---|
| $1d_{5/2}$ | 3.493 | | 3.211 | |
| $1f_{5/2}$ | 4.515 | | 4.083 | |

| | | |
|---|---|---|
| $1f_{7/2}$ | 4.489 | 4.061 |
| $1g_{7/2}$ | 5.519 | 4.986 |
| $1g_{9/2}$ | 5.485 | 4.955 |

**(b) Wave Functions:** $G(r) = m_1 \int \sqrt{|U(r)|}\, dr = \sqrt{\frac{2m}{\hbar^2}} \int \sqrt{U(r)}\, dr = Q(r)$

$Q(r) =$
$\sqrt{\frac{2m}{\hbar^2}} \Big\{ \frac{1}{2}\sqrt{a r^4 - C_{\ell sj} r^2 + b} -$
$\frac{C_{\ell sj}}{4\sqrt{a}} \ln\left[ 2\sqrt{a r^4 - C_{\ell sj} r^2 + b} + \frac{2 a r^2 - C_{\ell sj}}{\sqrt{a}} \right] - \frac{\sqrt{b}}{2} \ln\left[ 2\sqrt{b(a r^4 - C_{\ell sj} r^2 + b)} - C_{\ell sj} r^2 + 2b \right] +$
$\sqrt{b} \ln(r) \Big\}$

$$F(r) = A \cos[K r]\, e^{i G(r)} = A \cos[K r]\, e^{i Q(r)} = A \cos\left[\frac{q}{d} r\right] e^{i Q(r)} \quad (5a)$$

$$F(r) = B \sin[K r]\, e^{i G(r)} = B \sin[K r]\, e^{i Q(r)} = \sin\left[\frac{q}{d} r\right] e^{i Q(r)} \quad (5b)$$

$$|A| = |B| = \sqrt{2/d} = \sqrt{2 K/q} = \sqrt{q/d}\,;\quad \Psi(r,\theta,\phi) = R(r)\, Y(\theta,\phi) = \frac{F(r)}{r} Y(\theta,\phi) \quad (6)$$

The known radial wave functions are [20]: $R_{n\ell}(\rho) = A\, \rho^{\ell+1} e^{-\frac{1}{2}\rho^2} \mathcal{L}_n^{\ell+1/2}(\rho^2)$ \quad (7)

Here $\rho = \sqrt{m\omega/\hbar}\, r$ and $\mathcal{L}$ is the Laguerre polynomial.

*6.5. Three axial deformed harmonic oscillator potential (Nilsson model in the nuclear physics)*
It has been examined in detail at the source [36].

*6.6. Periodic potential of arbitrary form*
It has been examined in detail at the source [35] with the examples.

*6.7. Trigonometric and Hyperbolic Pöschl-Teller potential wells*
It has been examined in detail at the source [34].

*6.8. Saxon-Woods type central potential*
Saxon-Woods type potential function is given as: $V_1(r) = -\frac{V_0}{1+e^{[(r-R_0)/a]}}$ ; [a, $V_0$, $R_0$ parameters].
The spin-orbit interaction potential energy term can be added to this potential. The spin-orbit interaction potential energy term is given as follows: $V_{\ell sj}(r) = -\frac{\hbar^2}{4\mu^2 c^2} \frac{1}{r} \frac{dV_1(r)}{dr} [j(j+1) - \ell(\ell+1) - s(s+1)]$,
[$\mu$ is reduced mass ]. If this function is calculated, the following potential is found:

$$V_{\ell sj} = -\frac{\hbar^2 c^2 V_0\, e^{-R_0/a}}{4\mu^2 c^4\, a} [j(j+1) - \ell(\ell+1) - s(s+1)] \frac{1}{r} \frac{e^{r/a}}{\left[1+e^{\left[\frac{r-R_0}{a}\right]}\right]^2} = -C_{\ell sj} \frac{1}{r} \frac{e^{r/a}}{\left[1+e^{\left[\frac{r-R_0}{a}\right]}\right]^2}$$

$$C_{\ell sj} = \frac{\hbar^2 c^2 V_0\, e^{-R_0/a}}{4\,\mu^2 c^4\, a}\,[j(j+1) - \ell(\ell+1) - s(s+1)]$$

The centrifugal energy: $V_c(r) = \frac{\hbar^2 \ell(\ell+1)}{2\,\mu}\frac{1}{r^2}$. You must also get the Coulomb potential $V_{cp}(r)$ for charged particles if they exist. So the effective potential can be obviously written as:

$$U(r) = V_1(r) + V_{\ell sj} + V_c(r) + V_{cp}(r) = -\frac{V_0}{1+e^{\left[\frac{r-R_0}{a}\right]}} - C_{\ell sj}\frac{1}{r}\frac{e^{r/a}}{\left[1+e^{\left[\frac{r-R_0}{a}\right]}\right]^2} + \frac{\hbar^2 \ell(\ell+1)}{2\,\mu}\frac{1}{r^2} + V_{cp}(r) \quad (1)$$

The classical turning points of this effective potential cannot be found analytically, therefore, it must be found numerically. The classical turning points are the roots of the equation, $U(r) = -|E_q| = -M_h\,q^2/y$. Let these roots be $r_1(y)$ and $r_2(y)$. With these roots, first it is found: $d^2 = [r_2(y) - r_1(y)]^2 = d_2(y)$, after solving the equation of $d_2(y) = y$, it is found $y$ and $E_q = M_h\,q^2/y$.

**Numeric calculations:** Let us calculate energy values of $Cu(29, 68)$. We have taken the potentials as follows: $V_{cp}(r) = (Z-1)e^2\,\frac{3\,R_{co}^2 - r^2}{2\,R_{co}^3}$, [Coulomb potential in the sphere]

$V(r) = -\frac{V_0}{1+e^{(r-R_0)/a_0}}$ , (Saxon − Woods potentials)

$\alpha(L, S, J) = 0.5\,[J(J+1) - L(L+1) - S(S+1)]$

$V_{LSJ}(r) = -\frac{\hbar^2}{2\,\mu}\frac{1}{r}\,\alpha(L, S, J)\,\frac{V_{so}}{a_{so}}\frac{e^{(r-R_{so})/a_{so}}}{[1+e^{(r-R_{so})/a_{so}}]^2}$ , (spin − orbit potential)

For protons: $U(r) = V(r) + V_{LSJ}(r) + V_c(r) + V_{cp}(r)$; for neutrons: $U(r) = V(r) + V_{LSJ}(r) + V_c(r)$.

The parameters in these potentials have been calculated by the method of Volya [21, 22] and their values are: $a_{so} = a_0 = 0.662$; $R_{co} = R_0 = 5.142885$; $V_0 = 47.655271$; $V_{so} = 28.422020$;

$R_{so} = 4.726557$. The values of $d$ are obtained by solving the following equation,

$U(-d/2) + U(d/2) = -2\,\frac{\hbar^2}{2\,\mu}\frac{q^2}{d^2}$ and energy values have been calculated with the following formula:

$E_q = -\frac{\hbar^2}{2\,\mu}\frac{q^2}{d^2}$. The results are seen in the table 4.

**Table 4.** A few energy values of the Cu (29, 68) nucleus with Saxon-Woods potential (unit MeV)

| States | $E_q$(MeV) (neutron) | $E_q$ (MeV) (proton) |
|---|---|---|
| 1 s 1/2 | -45.6256 | -33.9536 |
| 1 p 3/2 | -15.1995 | -11.3653 |
| 1 p 1/2 | -15.1987 | -11.3644 |
| 1 d 5/2 | -6.5021 | -4.9028 |
| 1 d 3/2 | -6.5011 | -4.9015 |
| 2 s 1/2 | -45.6061 | -33.9536 |

| 1 f 7/2 | -3.4830 | -2.6484 |
| 2 p 3/2 | -25.1702 | -18.8825 |

*6.9. Relativistic Dirac equation in a central potential*

Consider a Dirac particle (spin is $1/2$) of mass m captured in a central potential well, V(r). With this potential, Dirac Hamiltonian can be written as follows[8, 9]: $H_D = \vec{\alpha}.\vec{p} + \beta m + V(r)$. Here, we have relativistic units: $\hbar = c = 1$. $\vec{p} = -i \vec{\nabla}$; $\vec{\alpha} = (\alpha_x, \alpha_y, \alpha_z)$ and $\beta$ are hermitical 4-operators acting on the spin variables alone. Including the position vector $\vec{r}$; $\vec{\alpha} = d\vec{r}/dt$ (velocity). The Hamiltonian $H_D$ and radial equation of Schrödinger can be brought to the form below [23]:

$$\frac{d^2 F(r)}{dr^2} + [\alpha - U(r)]F(r) = 0 \; ; \; \alpha = E^2 - m^2 \; ; \; U(r) = U_{re}(r) + U_j(r) + U_L(r) \pm U_{rez}(r) \quad (1)$$

$$U_{re}(r) = 2 E V(r) - V(r)^2 \; ; \; U_j(r) = (j + 1/2)^2/r^2 \; ; \; U_L(r) = L(L + 1)/r^2 \; ;$$

$$U_{rez}(r) = \sqrt{(j + 1/2)^2/r^4 - V'(r)^2}$$

For $\alpha > U(r)$ bound state: $k = i \sqrt{\alpha}$ ; $G(r) = \int \sqrt{U(r)} \, dr$. $r_1$ and $r_2$ are the roots of the equation:

$$[\alpha - U(r)] = 0 \text{ or } (E^2 - m^2) - U(r) = 0 \text{ and } d = r_2 - r_1 \; ; \; (r_1 < r_2 ).$$

The bound state energies are given by the solution of the equation:

$$K d = \sqrt{|\alpha|} \, d = d \sqrt{|E^2 - m^2|} = d \sqrt{m^2 - E^2} = q \; , \; [m > |E| \,] \quad (2)$$

We have; for $q = 2$ the ground state; for $q = n \pi$, $(n = 1, 2, 3, ...)$ the excited states.

Now let us find the function F(r) in bound states. In bound states, always $E > U(r)$, that is $\alpha > U(r)$.

Therefore: $K = \sqrt{|\alpha|} = \sqrt{|E^2 - m^2|} = \sqrt{m^2 - E^2} = K > 0$

$$G(r) = i \int \sqrt{U(r)} \, dr = i \int \sqrt{-|U(r)|} \, dr = Q(r)$$

$$F(r) = A \cos[K r] \, e^{i G(r)} = A \cos[K r] \, e^{i Q(r)} = A \cos\left[\frac{q}{d} r\right] e^{i Q(r)} \quad (3a)$$

$$F(r) = B \sin[K r] \, e^{i G(r)} = B \sin[K r] \, e^{i Q(r)} = \sin\left[\frac{q}{d} r\right] e^{i Q(r)} \quad (3b)$$

The application to the atoms of hydrogen-like has been given in detail in reference [23].

## 7. Transmission coefficient for an arbitrary form potential barrier

*7.1. Determination of the wave functions*

Let us consider the solution as follows: $F(r) = A \, e^{k r \pm i G(r)} + B \, e^{-k r \mp i G(r)}$ \quad (7.1)

Here, **(a)** For the case of $E > U(r)$, $k = i m_1 \sqrt{E}$, $G(r) = i m_1 \int \sqrt{U(r)} \, dr$

 **(b)** For the case of $E < U(r)$, $k = m_1 \sqrt{E}$, $G(r) = m_1 \int \sqrt{U(r)} \, dr$

Let us divide the potential into three domains, as seen in figure 3. In the region I, $E > U_1$; in the region II, $E < U_2$ and in the region III, $E > U_3$. Now, consider that a particle with total energy E comes from the left as in figure 3 and hits the barrier at the point $r_1$.

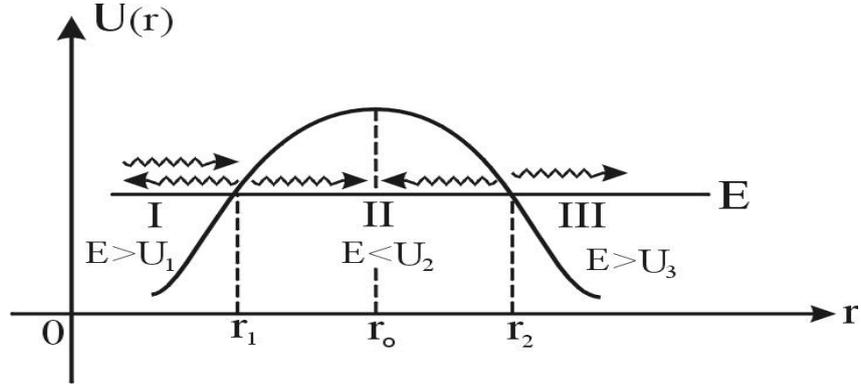

**Figure 3.** The unbounded state (potential barrier) and tunneling

According to the function (7.1), the wave functions can be obtained for these three regions as follows:

$$F_1(r) = A_1 e^{i K r \pm Q_1(r)} + B_1 e^{-i K r \mp Q_1(r)}$$
$$F_2(r) = A_2 e^{K r \mp i Q_2(r)} + B_2 e^{-K r \pm i Q_2(r)} \qquad (7.2)$$
$$F_3(r) = A_3 e^{i K r \pm Q_3(r)}$$

Here, $K = m_1\sqrt{E}$, $Q_p(r) = m_1 \int \sqrt{U_p(r)}\, dr$, $(p = 1, 2, 3)$

In the calculations of the above functions, the fact that the waves travel both from left to right and right to left in the region I and only from left to right in the region III have been taken into account. Since there is no wave coming from right to left in the region III, so the coefficient $B_3$ must be zero.

### 7.2. Calculation of the transmission coefficient T

To calculate the transmission coefficient T, the coefficients $A_i$ and $B_i$ in the functions given by (7.2) must be found. In order to find these coefficients, the following boundary conditions are used:

$$F_1(r_1) = F_2(r_1) \;;\; F_1'(r_1) = F_2'(r_1) \;;\; F_2(r_2) = F_3(r_2) \;;\; F_2'(r_2) = F_3'(r_2)$$

$$E = U_1(r_1) = U_2(r_1) \;;\; E = U_2(r_2) = U_3(r_2) \;;\; m_1\sqrt{E} = m_1\sqrt{U_p(r_1)} = m_1\sqrt{U_p(r_2)} = K$$

$$Q_p(r) = m_1 \int \sqrt{U_p(r)}\, dr \;\to\; Q_p'(r) = m_1\sqrt{U_p(r)} \;;\; Q_p'(r_1) = Q_p'(r_2) = K$$

In the functions (7.2), there are five unknown coefficients. Four of them can be found in term of $A_1$. All of the coefficients have been calculated in this study, but here, only the coefficient $A_3$ is sufficient. According to the signs of the exponential terms in (7.2), two expressions for $A_3$ can be found.

For the lower part of signs: $A_3 = \dfrac{2\exp[(1+i)K r_1 + (1-i)K r_2 - Q_1(r_1) + i Q_2(r_1) + i Q_2(r_2) + Q_3(r_2)]}{\exp[2 K r_1 + 2 i Q_2(r_1)] + \exp[2 K r_2 + 2 i Q_2(r_2)]} A_1$

For the upper part of signs: $A_3 = \dfrac{2\exp[(1+i)K r_1 + (1-i)K r_2 + Q_1(r_1) + i Q_2(r_1) + i Q_2(r_2) - Q_3(r_2)]}{\exp[2 K r_1 + 2 i Q_2(r_2)] + \exp[2 K r_2 + 2 i Q_2(r_1)]} A_1$

The transmission coefficient is defined as:

$$T = \frac{A_3 A_3^*}{A_1 A_1^*} = \frac{2}{\cosh[2 K d] + \cos[2 P]} \;;\; \left[P = Q_2(r_2) - Q_2(r_1) = \sqrt{\frac{2m}{\hbar^2}} \int_{r_1}^{r_2} \sqrt{U_2(r)}\, dr\,\right] \qquad (7.3)$$

Here, $Q_2(r) = m_1 \int \sqrt{U_2(r)}\, dr = \sqrt{\dfrac{2m}{\hbar^2}} \int \sqrt{U_2(r)}\, dr$

From the literature [6], the transmission coefficient T(or the barrier penetration probability) which is calculated by the method WKB is known as: $T = e^{-2g}$, $[g = \sqrt{\frac{2m}{\hbar^2}} \int_{r_1}^{r_2} \sqrt{U_2(r) - E} \, dr ]$ (7.4)

In (7.3) and (7.4), $r_1$ and $r_2$ are abscises of the points that the particle hits and leaves the potential barrier, respectively.

*7.3. Application to cold emission*
*7.3.1. Calculation of transmission coefficient.*
The cold emission of electrons from a metal surface is the basis of an important device known as a scanning tunneling microscope, or an STM. An STM consists of a very sharp conducting probe which is scanned over the surface of a metal (or any other solid conducting medium). A large voltage difference is applied between the probe and the surface. The surface electric field-strength immediately below the probe tip is proportional to the applied potential difference, and inversely proportional to the spacing between the tip and the surface. Electrons tunneling between the surface and the probe tip give rise to a weak electric current. The magnitude of this current is proportional to the tunneling probability T. It follows that the current is an **extremely sensitive** function of the surface electric field-strength, and, hence, of the spacing between the tip and the surface (assuming that the potential difference is held constant). An STM can thus be used to construct a very accurate contour map of the surface under investigation. In fact, STMs are capable of achieving sufficient resolution to image individual atoms.

Suppose that a cold metal surface is subject to a large uniform external electric field of strength $\epsilon$, which is directed such that it accelerates electrons away from the surface. The electrons just below the surface of a metal can be regarded as being in a potential well of depth W, where W is called the work function of the surface. Adopting a simple one-dimensional treatment of the problem let the metal lie at $x < 0$, and the surface at $x = 0$. The applied electric field is shielded from the interior of the metal. Hence, the energy E, say, of an electron just below the surface is unaffected by the field. In the absence of the electric field, the potential barrier just above the surface is simply $U(x) - E = W$. The electric field modifies this to $U(x) - E = W - e \epsilon x$. The potential barrier is sketched in figure 4.

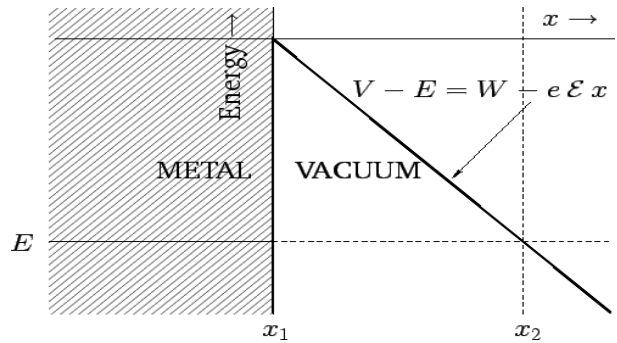

**Figure 4.** The potential barrier for an electron in a metal surface subject to an external electric field

It can be seen, from figure 4, that an electron just below the surface of the metal is confined by a triangular potential barrier which extends from $x = x_1$ to $x_2$, where $x_1 = 0$ and $x_2 = W/(e\,\epsilon)$. In the equation (7.3), if it is put that:

$$d = x_2 - x_1 = W/(e\,\epsilon)\;;\; m_1 = \sqrt{\frac{2m}{\hbar^2}}\;;\; K = \sqrt{\frac{2m}{\hbar^2}E} = m_1\sqrt{E} = m_1\sqrt{-|E|} = I\,m_1\sqrt{W}\;;$$

$$Q_2(x) = m_1 \int \sqrt{-e\,\epsilon\,x} = i\,\frac{2\sqrt{2\,m\,e\,\epsilon}}{3\,\hbar}x^{3/2} = i\sqrt{\frac{8\,m\,e\,\epsilon}{9\,\hbar^2}}\,x^{3/2}$$

It is obtained the following transmission coefficient:

$$T = T_{new} = \frac{2}{\cosh[2\,K\,d] + \cos[2\,Q_2(x_2)]} = \frac{2}{\cos\left[\frac{2\sqrt{2\,m}}{\hbar\,e\,\epsilon}W^{3/2}\right] + \cosh\left[\frac{4\sqrt{2\,m}}{3\,\hbar\,e\,\epsilon}W^{3/2}\right]} \quad (1)$$

Here, in calculation of (1), it is used that: $\cosh(i\,y) = \cos(y)$ and $\cos(i\,y) = \cosh(y)$

Making use of the WKB approximation [6], the probability of such an electron tunneling through the barrier and consequently being emitted from the surface is

$$T_{wkb} = \exp\left[-2m_1 \int_{x_1}^{x_2}\sqrt{U(x) - E}\,dx\right] = \exp\left[-2m_1 \int_{x_1}^{x_2}\sqrt{W - e\,\epsilon\,x}\,dx\right] = \exp\left[-\frac{4\sqrt{2\,m}}{3\,\hbar\,e\,\epsilon}W^{3/2}\right]$$

$$T_{wkb} = \exp\left[-\frac{4\sqrt{2\,m}}{3\,\hbar\,e\,\epsilon}W^{3/2}\right] \quad (2)$$

The above result (2) is known as the Fowler-Northeim formula. This formula is the result of WKB approximation. The formula (1) is exact formula, because there is not any approximation.

### 7.3.2. *Numerical calculations and comparison to the (1) and (2) formulas*

The barrier penetration probabilities or the transmission coefficients T have been calculated from the equations (1) and (2). In the calculations, the electron mass, $m\,c^2 = 0.511003$ MeV; the electron charge, $e = 1.602189 * 10^{-19}$ C $= 1.19999$ (MeV.fm)$^{1/2}$ have been taken. The obtained values for the different metals are seen in table 5. From the table 5, it can be seen that the new method is more appropriate than classical method WKB. In the calculations of the current-voltage characteristics of a diode in semiconductor physics, it is expected to have better results. [$\hbar\,c = 197.329$ MeV.fm]

**Table 5**. Comparison of the transmission coefficients values calculated with classical and new method

| Metals | Work function W (eV) [4] | Transmission coefficient $T_{new}$ From equation (1) | | | Transmission coefficient $T_{wkb}$ From equation (2) | | |
|---|---|---|---|---|---|---|---|
| | | Electric field $\epsilon$ (V/cm) | | | Electric field $\epsilon$ (V/cm) | | |
| | | $5\times10^6$ | $5\times10^7$ | $1\times10^7$ | $5\times10^6$ | $5\times10^7$ | $1\times10^7$ |
| Na | 2.46 | $5.12448*10^{-23}$ | 0.0206 | $1.43171*10^{-11}$ | $1.28112*10^{-23}$ | 0.0051 | $3.57928*10^{-12}$ |
| Al | 4.08 | $5.07471*10^{-49}$ | 0.000052 | $1.42474*10^{-24}$ | $1.26868*10^{-49}$ | 0.000013 | $3.56185*10^{-25}$ |
| Cu | 4.70 | 1.40147*10⁻⁶⁰ | 3.60174*10⁻⁶ | 2.36768*10⁻³⁰ | 3.50368*10⁻⁶¹ | 9.00435*10⁻⁷ | 5.91919*10⁻³¹ |
| Zn | 4.31 | 3.25862*10⁻⁵³ | 0.000020 | 1.14169*10⁻²⁶ | 8.14656*10⁻⁵⁴ | 4.91018*10⁻⁶ | 2.85422*10⁻²⁷ |
| Ag | 4.73 | 3.68836*10⁻⁶¹ | 3.15165*10⁻⁶ | 1.21464*10⁻³⁰ | 9.22090*10⁻⁶² | 7.87911*10⁻⁷ | 3.03659*10⁻³¹ |
| Pt | 6.35 | 4.59212*10⁻⁹⁵ | 1.28249*10⁻⁹ | 1.35530*10⁻⁴⁷ | 1.14803*10⁻⁹⁵ | 3.20624*10⁻¹⁰ | 3.38826*10⁻⁴⁸ |
| Pb | 4.14 | 4.19617*10⁻⁵⁰ | 0.000040 | 4.09691*10⁻²⁵ | 1.04904*10⁻⁵⁰ | 0.000010 | 1.02423*10⁻²⁵ |
| Fe | 4.50 | 9.20656*10⁻⁵⁷ | 8.67486*10⁻⁶ | 1.91902*10⁻²⁸ | 2.30164*10⁻⁵⁷ | 2.16872*10⁻⁶ | 4.79754*10⁻²⁹ |

*7.4. Application to alpha decay in atom nuclei and calculation of half-life*
*7.4.1. Calculation of half-life formula*

An $\alpha$ −particle is the nucleus of a helium atom. It consists of two protons and two neutrons. In the process of $\alpha$ − decay of nuclei, an α-particle is assumed to move in a spherical region determined by the daughter nucleus. The central feature of this one-body model is that the α-particle is preformed inside the parent nucleus. The success of the theory does not prove that $\alpha$-particle is preformed but that it behaves as if it was [24]. Figure 5 shows a plot, suitable for purposes of the theory, of the potential energy between the α-particle and the residual nucleus for various distances between their centers. The horizontal line $E_\alpha$ is the disintegration energy. There are three regions of interest. In the spherical region $r < r_1$, we are inside the nucleus and speak of a potential well with of depth $-U_0$, where $U_0$ is taken as a positive number. Classically, the α-particle can move in this region with a kinetic energy $E_\alpha + U_0$ but cannot escape from it. The region $r_1 < r < r_2$ forms a potential barrier because here the potential energy is more than the total available energy $E_\alpha$. The region $r > r_2$ is a classically permitted region outside the barrier. From the classical point of view, an α-particle in the spherical potential well would reverse its motion every time it tried to pass beyond $r = r_1$ of tunneling through such a barrier. A consistent model for this process assumes that an α-particle is bounded to the nucleus by a spherical potential well $V_1(r)$, or a spherical effective potential well $U_1(r)$, and that the α-particle is repelled from the residual nucleus by the central Coulomb potential barrier $V_2(r)$, or the effective central Coulomb potential barrier $U_2(r)$. The original radioactive nucleus has the charge $Z e$ and the α-particle has the charge $2e$. So the Coulomb potential barrier is as follows,

$$V_2(r) = \frac{2 (Z-2)e^2}{r} = \frac{c}{r}, \quad [c = 2 (Z-2)e^2] \tag{1}$$

Thus, the corresponding effective potential function is obtained as,

$$U_2(r) = \frac{c}{r} + \frac{b}{r^2}, \quad [b = \frac{\hbar^2}{2 m} \ell(\ell+1)] \tag{2}$$

This effective potential is depicted in figure 5. There are three domains in this effective potential.

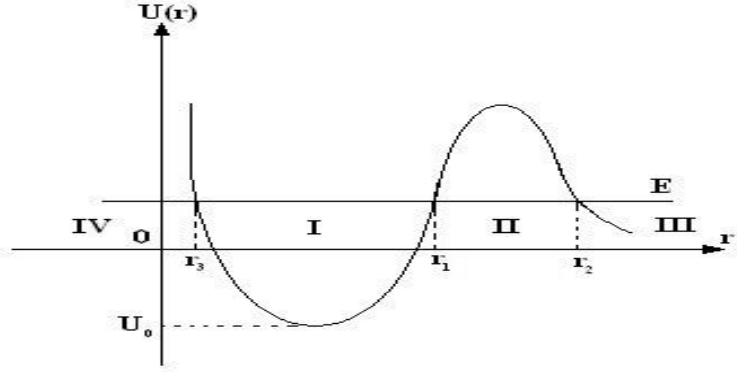

**Figure 5**. Effective potential function for the α-decay process of the nuclei

According to the one-body theory, the disintegration constant λ of an alpha emitter is given by:

$$\lambda = fT, \quad [\text{T transmission coefficient}] \tag{3}$$

Here f is the frequency with which the α-particle presents itself at the barrier and T is the probability of transmission through the barrier. The quantity f is roughly of the order of $\frac{v}{2(r_1-r_3)}$, where v is relative speed of the alpha particle inside the nucleus. $t_{1/2} = 0.693/\lambda$ is used for the calculations of nuclear half-life. The relative speed of the alpha particle can be found from its kinetic energy that is equal to the difference between the disintegration energy of the alpha particle, $E_\alpha$ and the ground state energy of the nucleus, $E_0$, namely,

$$\text{K.E.} = \frac{1}{2} m v^2 = [E_\alpha - (-|E_0|)] = E_\alpha + |E_0| \tag{4}$$

From the equation (4) it is obtained: $v = \sqrt{\frac{2(E_\alpha+|E_0|)}{m}}$ and $\frac{1}{v} = \sqrt{\frac{m}{2(E_\alpha+|E_0|)}}$ (5)

If the above values of (5) and (3) are substituted into $t_{1/2} = 0.693/\lambda$, then the following is obtained:

$$t_{1/2} = \frac{0.693}{\lambda} = \frac{0.693}{fT} = 0.693 \frac{2(r_1-r_3)}{T} \frac{1}{v} = 0.693 \frac{2(r_1-r_3)}{T} \sqrt{\frac{m}{2(E_\alpha+|E_0|)}} \tag{6}$$

To simplify the numerical calculations, the equation (6) can be rewritten as follows:

$$t_{1/2} = 0.693 \frac{2(r_1-r_3)}{c} \sqrt{\frac{mc^2}{2(E_\alpha+|E_0|)}} \frac{1}{T} \tag{7}$$

Here, $m = \frac{m_n m_\alpha}{m_n + m_\alpha}$ is the reduced mass. ($m_n$ and $m_\alpha$ are the mass of the nucleus and α−particle).

*7.4.2. Determination of the potential functions.*

In our numerical calculations, we have taken the harmonic oscillator type central potential as seen in figure 6. The central potential parts $V_1(r), V_2(r), V_3(r)$ and the central effective potential parts $U_1(r), U_2(r), U_3(r)$ are as follows:

$$V_1(r) = -U_0 + a\,r^2, \quad U_1(r) = -U_0 + a\,r^2 + \frac{b}{r^2} \ ; \quad [r_3 \leq r \leq r_1] \qquad \text{[I region]}$$

$$V_{21}(r) = a\,r^2, \quad U_{21}(r) = a\,r^2 + \frac{b}{r^2} \ ; \quad [r_1 \leq r \leq r_m] \qquad (\text{II}_1 \text{ region})$$

$$V_{22}(r) = \frac{c}{r}, \quad U_{22}(r) = \frac{c}{r} + \frac{b}{r^2} \ ; \quad [r_m \leq r \leq r_2] \qquad (\text{II}_2 \text{ region})$$

$$V_3(r) = \frac{c}{r}, \quad U_3(r) = \frac{c}{r} + \frac{b}{r^2} \ ; \quad [r_2 \leq r \leq \infty] \qquad (\text{III region})$$

Here, $R = R_0[(A-4)^{1/3} + 4^{1/3}]$ is the total rayon of the nucleus and alpha particle. From the solution of the equation $E_\alpha = U_1(r) = -U_0 + a\,r^2 + \frac{b}{r^2}$, we can obtain:

$$r_1 = \sqrt{\frac{(E_\alpha+U_0)+\sqrt{(E_\alpha+U_0)^2-4\,a\,b}}{2\,a}} \ ; \quad r_3 = \sqrt{\frac{(E_\alpha+U_0)-\sqrt{(E_\alpha+U_0)^2-4\,a\,b}}{2\,a}}$$

From the solution of the equation $E_\alpha = U_{22}(r) = \frac{c}{r} + \frac{b}{r^2}$ we can obtain: $r_2 = \frac{c+\sqrt{c^2+4\,b\,E_\alpha}}{2\,E_\alpha}$

From the solution of the equation $E_0 = U_1(r) = -U_0 + a\,r^2 + \frac{b}{r^2}$ we can obtain:

$$r_{11} = \sqrt{\frac{(E_0+U_0)-\sqrt{(E_0+U_0)^2-4\,a\,b}}{2\,a}} \ ; \quad r_{12} = \sqrt{\frac{(E_0+U_0)+\sqrt{(E_0+U_0)^2-4\,a\,b}}{2\,a}} \ ; \quad d_0 = r_{12} - r_{11}; \ d = r_2 - r_1$$

From the solution of the equation $E_0 = -\frac{2\,\hbar^2}{m\,d_0^2}$, we obtain the ground state energy as follows:

$$E_0 = \sqrt{a\,b} - \frac{1}{2}\left[U_0 + \sqrt{4\,a\,b - \frac{8\,a\,\hbar^2}{m} - 4\,\sqrt{a\,b}\,U_0 + U_0^2}\right]$$

With these grandeurs the coefficients of transmission are written as follow:

$$T_{wkb} = e^{-2\,P_{wkb}} \ ; \quad P_{wkb} = \sqrt{\frac{2\,m}{\hbar^2}}\left(\int_{r_1}^{r_m}\sqrt{U_{21}(r)-E_\alpha}\,dr + \int_{r_m}^{r_2}\sqrt{U_{22}(r)-E_\alpha}\,dr\right) \qquad (8)$$

$$T_{new} = \frac{2}{\cosh[2\,K\,d]+\cos[2\,\{Q_2(r_2)-Q_2(r_1)\}]} = \frac{2}{\cosh[2\,K\,d]+\cos[2\,P_{new}]} \qquad (9a)$$

$$P_{new} = \sqrt{\frac{2\,m}{\hbar^2}}\left(\int_{r_1}^{r_m}\sqrt{U_{21}(r)}\,dr + \int_{r_m}^{r_2}\sqrt{U_{22}(r)}\,dr\right) \qquad (9b)$$

$$t_{1/2}^{wkb} = 0.693\,\frac{2\,(r_1-r_3)}{c}\sqrt{\frac{m\,c^2}{2\,(E_\alpha-E_0)}}\,\frac{1}{T_{wkb}} \qquad (10)$$

$$t_{1/2}^{new} = 0.693\,\frac{2\,(r_1-r_3)}{c}\sqrt{\frac{m\,c^2}{2\,(E_\alpha-E_0)}}\,\frac{1}{T_{new}} \qquad (11)$$

The parameters $a$ and $r_m$ in potentials can be calculated. To calculate the parameter $a$, the following equation is used the equation: $U_1(r) = -U_0 + a\,r^2 + \frac{b}{r^2} = 0$. The following roots of this equation are found as: 
$$r_1' = \sqrt{\frac{U_0+\sqrt{U_0^2-4\,a\,b}}{2\,a}} \ ; \quad r_3' = \sqrt{\frac{U_0-\sqrt{U_0^2-4\,a\,b}}{2\,a}} \qquad (12)$$

As it can be seen in figure 6, the $r_1'$ can be taken as sum of the radii of the nucleus and the alpha particle. That is, since the radius of the alpha particle is $R_\alpha = R_0\,4^{1/3}$ and the radius of the nucleus $R_N = R_0(A-4)^{1/3}$, the total radius is: $R_c = R_\alpha + R_N = R_0[\,4^{1/3} + (A-4)^{1/3}]$ [24].

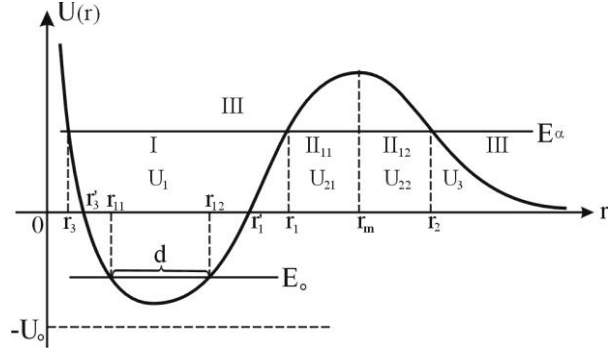

**Figure 6.** Central harmonic oscillator type effective potential

Thus it can be taken as follows: $(r_1')^2 = R_c^2 = [U_0 + \sqrt{U_0^2 - 4\,a\,b}]/(2\,a)$. From this equation, the value of $a$ is obtained as: $a = \dfrac{U_0}{R_c^2} - \dfrac{b}{R_c^4} = \dfrac{U_0 R_c^2 - b}{R_c^4}$ (13)

To calculate the parameter $r_m$, the equation $U_1(r) = U_{22}(r)$ is used and the resolution of this equation gives $r_m = \dfrac{2\sqrt[3]{3}\,a\,U_0 + \sqrt[3]{2}\,X}{6^{2/3}\,a\,X}$ ; $X = \left[9\,a^2 c + \sqrt{3\,a^3\,(27\,a\,c^2 - 4\,U_0^3)}\right]^{2/3}$ (14)

Therefore, the potential depends only on $R_0$ and $U_0$ parameters, which simplifies the calculations.

*7.4.3. Numerical calculations*

To calculate the transmission coefficient T (or barrier penetration probability) (8), (9) and the half-life values, the equations (10) and (11) have been used. Calculations were made for five nuclei and the results are visible on the tables 6 and 7. In these tables, the half-life values calculated from the new method and WKB method are compared with the experimental results. For the $R_0$ and $U_0$ parameters, first, $R_0 = 1.25$ fm and $U_0 = 40$ MeV values have been taken (table 6) and it was seen that $t_{1/2}^{new}$ values had better results than $t_{1/2}^{wkb}$ values. Later, $R_0$ values in the interval $1.10 - 1.60$ fm by step $0.01$ and $U_0$ values in the interval $30-50$ MeV by step 1 have been changed and the most consistent values with the experimental results have been taken (table 7). In all the tables, $I_i^{\pi_i}$ and $I_f^{\pi_f}$ are the initial and final state spins (parity) of the nucleus, respectively. $\ell_\alpha$ is the angular momentum of the alpha particle. From these tables, it can be seen that the new method is more appropriate than classical method (WKB). Table 7 shows that the new method gives very good results with the experiment [25-32]. In these tables "exp" is the experimental value of the half-life. y : year, d : day.

**Table 6.** Comparison of the half-life values of classical method with new method for $R_0 = 1.25$ fm and $U_0 = 40$ MeV in the harmonic oscilator potential model.

| $_Z^A X$ | $E_\alpha$ (MeV) | $I_i^{\pi_i}$ | $I_f^{\pi_f}$ | $\ell_\alpha$ | $t_{1/2}^{exp}$ | $t_{1/2}^{wkb}$ (3a) | $t_{1/2}^{new}$ (3b) | $t_{1/2}^{wkb}/t_{1/2}^{exp}$ | $t_{1/2}^{new}/t_{1/2}^{exp}$ |
|---|---|---|---|---|---|---|---|---|---|
| $_{84}^{208}$Po | 5.215 | $0^+$ | $0^+$ | 0 | 2.898 y | 0.0463 y | 5.6950 y | 0.16 | 1.97 |
| $_{84}^{210}$Po | 5.407 | $0^+$ | $0^+$ | 0 | 138.28 d | 1.2767 d | 267.865 d | 0.009 | 1.94 |

| $^{A}_{Z}X$ | $E_\alpha$ (MeV) | $I_i^{\pi_i}$ | $I_f^{\pi_f}$ | $\ell_\alpha$ | | | | | |
|---|---|---|---|---|---|---|---|---|---|
| $^{247}_{97}$Bk | 5.889 | $\frac{3}{2}^-$ | $\frac{5}{2}^-$ | 2 | 1380 y | 16.6389 y | 2521.68 y | 0.012 | 1.83 |
| $^{209}_{84}$Po | 4.979 | $\frac{1}{2}^-$ | $\frac{5}{2}^-$ | 2 | 102 y | 3.6377 y | 96.3285 y | 0.036 | 0.94 |
| $^{237}_{83}$Np | 4.959 | $\frac{5}{2}^+$ | $\frac{3}{2}^-$ | 1 | 2.14*10$^6$ y | 76826.3 y | 1.44*10$^6$ y | 0.036 | 0.67 |

**Table 7**. Comperason **of** the experimental half-life values and the results calculated by using the **new** methods with **the harmonic oscillator type well potential**, calculated values a and $r_m$ parameters for the parameters $r_0$ and $U_0$ which are very well matched with the experimental results.

| $^{A}_{Z}X$ | $E_\alpha$ (MeV) | $I_i^{\pi_i}$ | $I_f^{\pi_f}$ | $\ell_\alpha$ | $r_0$ (fm) | $U_0$ (MeV) | $t_{1/2}^{exp}$ | $t_{1/2}^{new}$ (3b) | $t_{1/2}^{new}/t_{1/2}^{exp}$ |
|---|---|---|---|---|---|---|---|---|---|
| $^{208}_{84}$Po | 5.215 | $0^+$ | $0^+$ | 0 | 1.30 | 45 | 2.898 y | 2.899 y | 1.0004 |
| $^{210}_{84}$Po | 5.407 | $0^+$ | $0^+$ | 0 | 1.30 | 47 | 138.40 d | 138.41 d | 1.0001 |
| $^{247}_{97}$Bk | 4.979 | $\frac{1}{2}^-$ | $\frac{5}{2}^-$ | 2 | 1.24 | 35 | 102 y | 102 y | 1.0010 |
| $^{209}_{84}$Po | 4.959 | $\frac{5}{2}^+$ | $\frac{3}{2}^-$ | 1 | 1.23 | 49 | 2.14*10$^6$ y | 2.15*10$^6$ y | 1.0028 |
| $^{237}_{83}$Np | 5.889 | $\frac{3}{2}^-$ | $\frac{5}{2}^-$ | 2 | 1.29 | 43 | 1380 y | 1375 y | 0.9964 |

Here, the general transmission coefficient formula for a potential barrier with an arbitrary form has been easily calculated without making any approximation. In this calculation, a new method that we developed for the solution of the radial SE has been used. The transmission coefficient obtained from the new method is given by the formula (7.3). In this formula, it could be difficult to calculate the integral $\int \sqrt{U_2(r)}\, dr$. If these calculations cannot be made analytically, it should then be performed by numerical methods.

In the application of the general transmission coefficient formula to the α-decay, three-dimensional harmonic oscillator potential well has been used. The results have been given in the tables together with the experimental values. The tables contain also the "ratio" column for comparison. It can be seen from table 7, the ratios for the most of the nuclei are close to 1 (one). The deviations from 1 are within the experimental error. Hence, it is said that the results obtained from the new method are more realistic. In the WKB method, the wave function is sinusoidal inside and outside the potential barrier, but is not sinusoidal and is an exponential function into the potential barrier. So the entering wave into the potential barrier is not sinusoidal and after the potential barrier it becomes again sinusoidal. But in new method, the wave function includes a sinusoidal multiplier inside the potential barrier, but the sinusoidal multipliers are different inside and outside the potential barrier. Thus, the wave function has different phases inside and outside the potential barrier, but it advances everywhere as sinusoidal functions. This can also be said that it is more accurate and realistic. Besides, the WKB method gives approximately a wave function. In the new method, the wave function is exact because there is no approximation. That is

H.H. Erbil

why the theoretical calculated half-life values match better with the experimental values. From these we conclude that the transmission coefficient given by the equation (7.3) is more correct and realistic. By using the new transmission coefficient and half-life formulas, the half-life values of nuclei can easily be calculated. The general transmission coefficient formula can be used for the other tunneling phenomenon, like the cold emission from the metals and so forth.

## 8. Scattering theory
*8.1. Calculation of the scattering amplitudes*

Let us consider a spherical wave progressing at the direction of $Oz$ axis from left to right, and arriving to a central potential field, sitting at the origin of the $Oxyz$ coordinate system. When we consider scattering, we shall assume that the interaction between the scattering particle and the scatter can be represented by an effective central potential energy function $U(r)$, here $r$ relative radial variable. The effective potential $U(r)$ can include the parts of attractive and repulsive. Such a central potential is schematically represented at figure 1. The total energy of the incoming particle beam is E and the incoming particle beam can be represented by the spherical wave. This progressive spherical wave progress from right to left and arrives to the point $r = r_1$ at the figure 7. We divide the potential region into four zones and examine the motion of particle beam into these four zones.

The zone I is the region before the effective potential from where free particle comes; the zone II, III and IV are the effective potential regions where the particle beam is affected. These regions may include attractive and repulsive potential segments. Zone IV is the region where the particle beam is not able to penetrate because of the potential is infinite. So the wave function is zero in this region. These are presented with four regions together with the central potentials in figure 7.

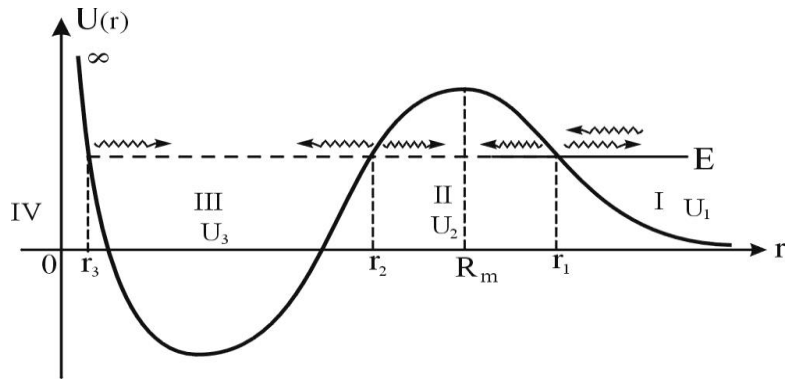

**Figure 7.** General schematic representation of scattering by central potential

We assume that $r = r_1$ and $r = r_2$ at the interface between the zone I and II, and the zone II and III, respectively. The effective potential segments in the zones are represented as $U_1(r)$, $U_2(r)$ and $U_3(r)$ according to the zone numbers. The effective potential $U_4(r)$ can be assumed infinite since the particle does not enter into this region. The central potential can be taken as zero at much far from the zone I, so that the particle is free in that region and the effective potential is composed of only the centrifugal term due to the incoming particle angular momentum or spin. The Coulomb interaction potential should also be added to $U_1(r)$ if that is available. The total energy of the incoming particle and the centrifugal term are always positive and the latter is less than the former. According to the functions (4.1), the following functions are determined for the zones that are taken into account:

In the zone I: $E > 0$, $U_1(r) > 0$ and $E > U_1(r)$; $k = i\,m_1\sqrt{E} = i\,K$; ($K = m_1\sqrt{E}$);

$$G_1(r) = i\,m_1 \int \sqrt{U_1(r)}\,dr = i\,Q_1(r)\,;\ \left[m_1 = \sqrt{\tfrac{2m}{\hbar^2}}\,,\ Q_1(r) = m_1 \int \sqrt{U_1(r)}\,dr\right] \quad (8.1a)$$

In the zone II: $E > 0$, $U_2(r) > 0$ and $E < U_2(r)$, $k = m_1\sqrt{E} = K$

$$G_2(r) = m_1 \int \sqrt{U_2(r)}\,dr = Q_2(r);\ \left[K = m_1\sqrt{E},\ m_1 = \sqrt{\tfrac{2m}{\hbar^2}}\,, Q_2(r) = m_1 \int \sqrt{U_2(r)}\,dr\right] \quad (8.1b)$$

In the zone III, the conditions $E < 0$ (this zone corresponds to the bound state, so negative energy should be taken), $U_3(r) < 0$ and $E > U_3(r)$ leads to the expressions below:

$$k = i\,m_1\sqrt{E} = i\,m_1\sqrt{-|E|} = \pm K,\ \left[K = m_1\sqrt{|E|}\right]$$

$$G_3(r) = i\,m_1 \int \sqrt{U_3(r)}\,dr = i\,m_1 \int \sqrt{-|U_3(r)|}\,dr = \pm\,Q_3(r),\left[Q_3(r) = m_1 \int \sqrt{|U_3(r)|}\,dr\right] (8.1c)$$

The wave function in the zone I should vanish at large distance. Under these circumstances, the radial wave functions in the three zones can be put in the forms below with regard to the general functions (4.1):

$$F_1(r) = A_1 e^{i\,K r - Q_1(r)} + e^{-i\,K r - Q_0(r)}\ ;\ [B_1 = 1,\ Q_1(r) > 0,\ Q_0(r) > 0] \quad (8.2a)$$

$$F_2(r) = A_2 e^{K r \pm i\,Q_2(r)} + B_2 e^{-K r \mp i\,Q_2(r)} \quad (8.2b)$$

$$F_3(r) = A_3 e^{-K r \pm i\,Q_3(r)} + B_3 e^{K r \mp i\,Q_3(r)} \quad (8.2c)$$

The wave function in the zone IV vanishes since the effective potential in this region is infinite. Here, $Q_0(r)$ is the function resulting from the angular momentum of the incoming particle. This function $Q_0(r)$, can also be taken as zero, because it has not any contribution to the calculation of the scattering cross-section, as it will be seen, soon.

The potential in the zone III can also be complex in some cases (usually called the optical potential). If $U_3(r)$ is the optical potential, it can be written as:

$$U_3(r) = |U_3(r)|e^{i\emptyset} = U_{31}(r) + i\,U_{32}(r) = \sqrt{U_{31}^2(r) + U_{32}^2(r)}\ e^{i\emptyset} \quad (8.3a)$$

Here, $\tan(\emptyset) = \dfrac{U_{32}(r)}{U_{31}(r)}$, $\emptyset = \arctan\left[\dfrac{U_{32}(r)}{U_{31}(r)}\right]$; $k = i\,m_1\sqrt{E} = i\,m_1\sqrt{-|E|} = \pm K,\left[K = m_1\sqrt{|E|}\right]$;

$$G_3(r) = i\,m_1 \int \sqrt{U_3(r)}\,dr = i\,m_1 \int \sqrt{-|U_3(r)|}\,dr = \pm\,Q_3(r)$$

$$Q_3(r) = m_1 \int \sqrt{|U_3(r)|}\,dr = m_1 \int \sqrt{\sqrt{U_{31}^2(r) + U_{32}^2(r)}}\,dr = m_1 \int \sqrt[4]{U_{31}^2(r) + U_{32}^2(r)}\,dr \quad (8.3b)$$

In the equations (8.2) and (8.3), the functions $Q_p(r)$ can be written briefly as follows:

$$Q_p(r) = m_1 \int \sqrt{|U_p(r)|}\,dr,\ [p = 0, 1, 2, 3] \quad (8.3c)$$

The terms containing $A_1$ and $B_1$ in the functions (8.2) give outgoing and incoming waves, respectively. We assume that the amplitude of incoming wave at the boundary of the zone I and II is constant. The second (8.2b) and the third (8.2c) functions represent the states of the wave in the effective region of the potential. Applying the continuity conditions on $F_p(r_j)$ and $F'_p(r_j)$, $[p, j = 1, 2, 3]$ functions, the coefficients $A_i$ and $B_i$ in the equations (8.2) can be determined. These conditions at the boundary points of the three zones can be written as in the following form:

$$F_1(r_1) = F_2(r_1)\,;\ F'_1(r_1) = F'_2(r_1)\,;\ F_2(r_2) = F_3(r_2)\,;\ F'_2(r_2) = F'_3(r_2) \quad (8.4)$$

$F_3(r_3) + F'_3(r_3)/K = 0$ , $Q'_p(r_j) = K$ , $(p, j = 1,2,3)$

The coefficients $A_1$, $A_2$, $B_2$, $A_3$, $B_3$ in the functions (8.2) can be found by solving five linear equations, which can be obtained by using the conditions (8.4) for each of the functions (8.2).

The essential coefficient for the scattering cross section is $A_1$, as described below. Therefore there is no need to give other coefficients here. The $A_1$ coefficient, which is obtained from four equations, is computed by taking into account the lower and upper signs in the exponential expressions in the equation set (8.2) as follows:

**The $A_1$ coefficient, found using the lower signs:**

$A_1 = \frac{1}{2}(1+i) \, e^{-2iKr_1 - Q_0(r_1) + Q_1(r_1)} \{-1 + (2-i) e^{[2K(r_3 - r_1) + 2i[Q_2(r_1) - Q_2(r_2) + Q_3(r_2) - Q_3(r_3)]]}\}$ (8.5a)

**The $A_1$ coefficient, found using the upper signs:**

$A_1 = \frac{(1+i) \, e^{2(1-i)Kr_1 - Q_0(r_1) + Q_1(r_1) + 2i[Q_2(r_1) + Q_3(r_2)]}}{(2+i) e^{[2Kr_3 + 2i[Q_2(r_2) + Q_3(r_3)]]} - e^{[2Kr_1 + 2i[Q_2(r_1) + Q_3(r_2)]]}}$ (8.5b)

The terms with $A_1$ in the functions (8.2) representing the outgoing wave from the center of potential, includes both the scattered by the potential and the incoming wave. Therefore, we must subtract away the latter to find the amplitude of only the scattered wave. Thus, we obtain the scattering amplitude and radial wave function representing the scattered wave as:

$C_s(r_1) = A_1 \, e^{-Q_1(r_1)} - e^{-Q_0(r_1)}$ , (scattering amplitude) (8.6)

$R_s(r) = \frac{F_s(r)}{r} = C_s(r_1) \frac{e^{iKr}}{r} = \left[A_1 \, e^{-Q_1(r_1)} - e^{-Q_0(r_1)}\right] \frac{e^{iKr}}{r}$ (8.7)

This function (8.7) represents the elastic scattering wave by the potential.

*8.2. Calculation of particle currents*
*8.2.1. Calculation of the scattered particle current*
Using the equation (8.7) and the equation: $J_s(r) = \frac{\hbar}{2mi} \left[R_s^*(r) \frac{dR_s(r)}{dr} - R_s(r) \frac{dR_s^*(r)}{dr}\right]$, the current of scattered particles per unit area at $r = r_1$ point can be found as:

$J_s(r_1) = \frac{1}{r_1^2} \frac{\hbar K}{m} |C_s(r_1)|^2$ (8.8)

*8.2.2. Calculation of incoming particle current.*
The radial wave function of the incoming beam, passing through the zone I and toward the zone II, can be written as: $R_g(r) = \frac{F_g(r)}{r} = \frac{1}{r} e^{-iKr - Q_0(r_1)} = C_g(r_1) \frac{e^{-iKr}}{r}$ , $[C_g(r_1) = e^{-Q_0(r_1)}]$

Here, $C_g(r_1) = e^{-Q_0(r_1)}$ represents the amplitude of incoming wave. The incident current per unite area at $r = r_1$ point can be obtained in the way that is applied to the current equation:

$J_g(r_1) = \frac{1}{r_1^2} \frac{\hbar K}{m} |C_g(r_1)|^2 = \frac{1}{r_1^2} \frac{\hbar K}{m} e^{-2Q_0(r_1)}$ (8.9)

*8.3. Calculations of scattering cross -sections*
*8.3.1. Calculation of elastic scattering differential cross section*
The probability per unit differential surface of a sphere of radius $r_1$, that an incident particle is scattered into the differential surface area on the sphere of radius $r_1$, $dS = r_1^2 \, d\Omega$, $[d\Omega = \sin(\theta) \, d\theta \, d\phi]$, is expressed as the ratio of the scattered current to the incident current, that is:

$$\frac{d\sigma_s}{dS} = \frac{d\sigma_s}{r_1^2 d\Omega} = \frac{J_s(r_1)}{J_g(r_1)} \quad \rightarrow \quad \frac{d\sigma_s}{d\Omega} = \frac{J_s(r_1)}{J_g(r_1)} r_1^2 \qquad (8.10)$$

The differential elastic cross section can be expressed in a simple form by putting (8.8) and (8.9) into the equation (8.10):

$$\frac{d\sigma_s}{d\Omega} = \frac{C_s(r_1) C_s^*(r_1)}{C_g(r_1) C_g^*(r_1)} r_1^2 = \frac{|C_s(r_1)|^2}{|C_g(r_1)|^2} r_1^2 \qquad (8.11)$$

Since the scattering is azimuthally symmetrical, the angle $\phi$ can be integrated out, so that the expression (8.10) and (8.11) can be written as follows:

$$\frac{d\sigma_s}{d\theta} = 2\pi \frac{J_s(r_1)}{J_g(r_1)} r_1^2 \sin(\theta) = 2\pi \frac{|C_s(r_1)|^2}{|C_g(r_1)|^2} r_1^2 \sin(\theta) \qquad (8.12)$$

The expressions (8.12) show the elastic scattering differential cross sections in the angle $d\theta$ which is usually measured experimentally.

*8.3.2. Calculation of differential inelastic or reaction (no-elastic) cross section*

Differential reaction (capture of particle, emission of particle, inelastic collision…) cross section per the solid angle can be found through the difference between the incoming current and the outgoing current divided by the former. By analogy with equation (8.12), the differential reaction cross section can be expressed as:

$$\frac{d\sigma_r}{d\theta} = 2\pi \frac{[J_g(r_1) - J_s(r_1)]}{J_g(r_1)} r_1^2 \sin(\theta) = 2\pi \frac{[|C_g(r_1)|^2 - |C_s(r_1)|^2]}{|C_g(r_1)|^2} r_1^2 \sin(\theta) \qquad (8.13)$$

*8.3.3. Calculation of total cross sections*

The total elastic scattering cross section is the total probability to be elastic scattered in any direction and it can be determined through the integral of differential cross section (8.11):

$$\sigma_s = \int d\sigma_s = \int \frac{d\sigma_s}{d\Omega} d\Omega = \iint \frac{J_s(r_1)}{J_g(r_1)} r_1^2 \sin(\theta) \, d\theta \, d\phi$$

$$\sigma_s = 4\pi r_1^2 \frac{J_s(r_1)}{J_g(r_1)} = 4\pi r_1^2 \frac{|C_s(r_1)|^2}{|C_g(r_1)|^2} \qquad (8.14)$$

By analogy with the equation (8.14), the total reaction cross section can be expressed as:

$$\sigma_r = 4\pi r_1^2 \frac{[J_g(r_1) - J_s(r_1)]}{J_g(r_1)} = 4\pi r_1^2 - \sigma_s \qquad (8.15)$$

In the equation (8.15), it is seen that: if $J_s(r_1) = J_g(r_1)$, then $\sigma_r = 0$, full-elastic scattering; if $J_s(r_1) > J_g(r_1)$, then $\sigma_r < 0$, that is taken out of the particle from the target (emission of particle from target) and if $J_s(r_1) < J_g(r_1)$, then $\sigma_r > 0$, that is captured (absorbed) the particle by the target.

The total scattering cross section, including all process [elastic plus reaction (all of no-elastic events)]:

$$\sigma_t = \sigma_s + \sigma_r = 4\pi r_1^2 \frac{J_s(r_1)}{J_g(r_1)} + 4\pi r_1^2 \frac{[J_g(r_1) - J_s(r_1)]}{J_g(r_1)} = 4\pi r_1^2 \qquad (8.16)$$

Then the cross-sections $\sigma_s$, $\sigma_r$, $\sigma_t$ can be expressed through the $A_1$ coefficients (8.5).

**(a)** $\sigma_s$ elastic scattering cross-section found using the coefficient (8.5a) is as follows:

$$X_1 = 3 e^{4 K r_1} + 5 e^{4 K r_3} + 2 e^{4 K r_1} [\cos(2 K r_1) + \sin(2 K r_1)]$$

$$X_2 = -2 e^{2 K (r_1 + r_3)} [3 \cos(2(K r_1 - Y)) + 2 \cos(2 Y) + \sin(2(K r_1 - Y)) + \sin(2 Y)$$

$$\frac{\sigma_s}{4 \pi r_1^2} = \frac{1}{2} e^{-4 K r_1} \{X_1 + X_2\} \qquad (8.17a)$$

**(b)** $\sigma_s$ elastic scattering cross-section found using the coefficient (8.5b) is as follows:

$$P_1 = 3\,e^{4\,K\,r_1} + 5\,e^{4\,K\,r_3} + 2\,e^{4\,K\,r_1}[\cos(2Kr_1) + \sin(2Kr_1)]$$

$$P_2 = -2\,e^{2\,K(r_1+r_3)}[3\cos(2Kr_1 - 2\,Y) + 2\cos(2\,Y) + \sin(2Kr_1 - 2\,Y) + \sin(2Y)]$$

$$P_3 = e^{4\,K\,r_1} + 5\,e^{4\,K\,r_3} - 2\,e^{2\,K(r_1+r_3)}[2\cos(2Y) + \sin(2Y)]$$

$$\frac{\sigma_s}{4\pi r_1^2} = \frac{P_1 + P_2}{P_3} \qquad (8.17b)$$

In both cases, the reaction and total scattering cross-section $\sigma_r$ and $\sigma_t$ are as:

$$\frac{\sigma_r}{4\pi r_1^2} = 1 - \frac{\sigma_s}{4\pi r_1^2} \quad \text{and} \quad \frac{\sigma_t}{4\pi r_1^2} = \frac{\sigma_s}{4\pi r_1^2} + \frac{\sigma_r}{4\pi r_1^2} = 1 \text{ or } \sigma_t = 4\pi r_1^2 \qquad (8.17c)$$

In these expressions, Y is given by:

$$Y = Q_2(r_1) - Q_2(r_2) + Q_3(r_2) - Q_3(r_3) = m_1 \int_{r_2}^{r_1} \sqrt{|U_2(r)|}\, dr + m_1 \int_{r_3}^{r_2} \sqrt{|U_3(r)|}\, dr \qquad (8.17d)$$

The integrals in the equation (8.17d) can be solved numerically if the functions $Q_2(r)$ and $Q_3(r)$ cannot be calculated analytically. It can be seen from these formulas (8.17) that the scattering cross-sections ($\sigma_s$ and $\sigma_r$) depend on the total energy E [with K(E)], Y integral and the effective radius $r_1$, $r_2$, $r_3$ of the scatter potential, separately, but total scattering cross-section, $\sigma_t = \sigma_s + \sigma_r$ only depends on the parameter $r_1$, so the energy E. Here, $r_1$ can be considered as impact or collision parameter, classically.

*8.4. Examples of the calculation of scattering cross-section*
*8.4.1. Model potentials, wave functions and their ingredients*

To calculate a scattering cross-section, a model potential should be considered. Here as an example, we consider the Wood-Saxon shape potential plus the spin-orbit, centrifugal and Coulomb potentials. The potential zones are defined in figure 1 and they are shown in figure 2 for these model potentials. Wood-Saxon potential depends on three parameters, ($V_0$, $a_c$, $R_c$). The calculations have been thus performed with these potentials.

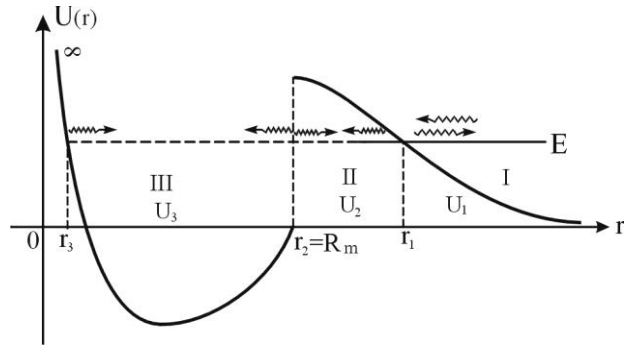

**Figure 8.** Effective Wood-Saxon potential and Coulomb potential zones used in the calculations

Scattering affects only relative motion. The scattering event affects only the relative motion. In the center of mass reference frame (CM), the scattering cross section of the incoming (incident or projectile) particle $\sigma(\Omega)$ depends on the energy $E_r = M_t E_L/(M_p + M_t)$, where $M_p$ and $M_t$, respectively, mass of incident (projectile) and target particles; $E_L$ Laboratory and $E_r$ relative energies. The boundary values of the potential zones, taken in the calculations: the maximum potential energy occurs at the distance $r_2 = R_m = R_0(A_p^{1/3} + A_t^{1/3})$, the addition of the projectile (incident particle) radius to the target radius, provided that they are spherical, here $A_p$ and $A_t$ are the mass number of the projectile and the target, respectively. The zones and $r_k$, $(k = 1, 2, 3)$ values are shown in figure 8. Effective potential energy is as:

$$U(r) = V_{ws}(r) + V_{LSJ}(r) + V_S(r) + V_c(r) \tag{8.18}$$

$$V_{ws}(r) = -\frac{V_0}{1+\text{Exp}[(r-R_c)/a_c]}, \text{ Wood-Saxon potential} \tag{8.19a}$$

$$V_{LSJ}(r) = -\frac{\hbar^2}{2 M_i^2 c^2} \frac{1}{r} \frac{dV(r)}{dr} <\vec{L}.\vec{S}>, \text{ spin-orbit potential} \tag{8.19b}$$

$$<\vec{L}.\vec{S}> = \frac{1}{2}[J(J+1) - L(L+1) - S(S+1)] \tag{8.19c}$$

$$V_S(r) = \frac{b}{r^2} \; ; \; [b = \frac{\hbar^2 L(L+1)}{2 M_i}], \text{ centrifugal potential,} \tag{8.19d}$$

$$V_c(r) = \frac{C_c}{r}, \; [C_c = (Z_p e)(Z_t e) = Z_p Z_t e^2], \text{ potential energy of Coulomb.} \tag{8.19e}$$

Here, $[L, S, J]$ are the relative orbital, spin and total angular momentum quantum numbers between the target and projectile, respectively. $M_i$ is the reduced mass of the target and projectile. $R_c = R_0 A_t^{1/3}$, ($R_0$ is parameter).

If $r_{sc} = [C_c + \sqrt{C_c^2 + 4 b E_r}]/[2 E_r]$ is the positive root of the equation $E_r = \frac{b}{r^2} + \frac{C_c}{r} = V_s(r) + V_c(r)$ and $r_2 = R_m$, then we get $r_1 = r_2 + r_{sc} = r_2 + [C_c + \sqrt{C_c^2 + 4 b E_r}]/[2 E_r]$. $Z_p$ and $Z_t$ are the charge numbers of the projectile and the target, respectively. Value $r_3$ is determined by equalizing $V_s(r)$ to $E_r$. Consequently, $(r_3, r_2, r_1)$ values are obtained as follows:

$$r_3 = \sqrt{b/E_r} \quad ; \quad r_2 = R_m = R_0\left(A_p^{\frac{1}{3}} + A_t^{\frac{1}{3}}\right) \; ; \; r_1 = r_2 + \frac{C_c + \sqrt{C_c^2 + 4 b E_r}}{2 E_r}$$

The wave functions for the zones 1-3, used in the calculation, can be expressed in the following ways, respectively.

$$Q_0(r) = m_\hbar \int \sqrt{\frac{\hbar^2 L_0(L_0+1)}{2 M_i r^2}} \; dr = \sqrt{L_0(L_0+1)} \; \ln(r) \tag{8.20a}$$

$$Q_1(r) = m_\hbar \int \sqrt{\frac{C_c}{r} + \frac{b}{r^2}} \; dr = 2 m_\hbar \left\{\sqrt{b + C_c r} - \sqrt{b} \; \text{arctanh}\left[\sqrt{\frac{b + C_c r}{b}}\right]\right\} \tag{8.20b}$$

$$Q_2(r) = m_\hbar \int \sqrt{\frac{C_c}{r} + \frac{b}{r^2}} \; dr = 2 m_\hbar \left\{\sqrt{b + C_c r} - \sqrt{b} \; \text{arctanh}\left[\sqrt{\frac{b + C_c r}{b}}\right]\right\} \tag{8.20c}$$

$$Q_3(r) = m_\hbar \int \sqrt{|U_3(r)|} \; dr = m_\hbar \int \sqrt{|V_{ws}(r) + V_{LSJ}(r) + V_S(r)|} \; dr \tag{8.20d}$$

Here, $m_\hbar = \sqrt{2M_i}/\hbar$ and $L_0$ is the angular momentum of the incident particle.

It is seen from the equation (8.17c) that, even though $\sigma_s$ and $\sigma_r$ have changed with the parameters $V_0$ and $a_c$; $\sigma_t = \sigma_s + \sigma_r$ has not changed for a certain value of $R_0$. In other words, the total cross-section does not depend on the parameters $V_0$ and $a_c$, therefore does not also depend on the potential. So, $R_0$ parameter can be obtained from the solution of the equation $4\pi r_1^2 = \sigma_t^{exp}$ as, ($\sigma_t^{exp}$ is experimental total cross-section):

$$R_0 = \frac{-\sqrt{\pi}\,10\left[C_c + \sqrt{C_c^2 + 4\,b\,E_r}\right] + E_r\sqrt{10\,\sigma_t^{exp}}}{20\,E_r\sqrt{\pi}\left[\sqrt[3]{A_p} + \sqrt[3]{A_t}\right]} \tag{8.21}$$

To calculate $\sigma_s$ and $\sigma_r$ separately, the parameter $V_0$ in the range 20-60 by step 0.0001 and $a_c$ in the range 0.40-0.60 by step 0.01 have been changed in the potential, until $\sigma_r^{cal} > 0$ and round $[\sigma_r^{exp}]$ =round $[\sigma_r^{cal}]$ and floor $[\sigma_r^{exp}]$ = floor $[\sigma_r^{cal}]$}, during the calculations. So, rough $V_0$ and $a_c$ values have been found. Then, with these approximate values, $\{\sigma_s^{exp} = \sigma_s^{cal}$ and $\sigma_r^{exp} = \sigma_r^{cal}\}$ system of equations has been solved and the exact values of parameters $V_0$ and $a_c$ have been found. With these $V_0$ and $a_c$ parameters, the exact $\sigma_s$ and $\sigma_r$ have been recalculated. We have taken in the calculations: $e^2 = 1.439976$ MeV fm ; $M_u = 931.502$ MeV $/c^2$ ; $\hbar c = 197.329$ MeV fm .

*8.4.2. Thermal neutron cross-sections*

For an example, thermal neutron ($E_L = 0.025$ eV) scattering cross-sections for some targets have been calculated and compared with the measured values. In calculation, because of their angular momentums are zero; even nuclei have been taken as targets. So the relative angular momentums are those of the projectile, they are $L = 0$, $S = 1/2$, $J = 1/2$ for neutron. Calculations of $\sigma_s$, $\sigma_r$ and $\sigma_t$, it has been used the formulas given by Eq. (8.17c) which has been obtained by the lower sign functions. The results are compared with the measured total cross-sections in the table 8. The experimental values have been taken from [25-32]. It is seen that agreements are fairly good as seen at the table 8. In this table: First column: target; Second column: ($R_0$, $V_0$, $a_c$) parameters; Third column: ($R_c$, $R_m = r_2$, $r_1$); Fourth column: cross-sections; Fifth column: values of calculation; Sixth column: values of experiment.

**Table 8.** [ n (0 , 1) + Xn (Z , N) ] Thermal neutron cross-section comparisons with those measured

| Target<br>Xn (Z, N) | $R_0$<br>$V_0$<br>$a_c$ | $R_c$<br>$R_m$<br>$r_1$ | Cross-sections<br>(mb) | Calculations<br>(mb) | Experiment<br>(mb) |
|---|---|---|---|---|---|
|  | 2.29845 | 2.89586 | $\sigma_s \rightarrow$ | 3390 | 3390±12 |
| H (1,2) | 21.3275 | 5.19431 | $\sigma_r \rightarrow$ | 0.519 | 0.519±0.007 |
|  | 0.40 | 5.19431 | $\sigma_t \rightarrow$ | 3390.52 | 3390.52 |
|  | 1.86896 | 4.27885 | $\sigma_s \rightarrow$ | 4746 | 4746±2 |
| C (6,12) | 35.2934 | 6.14781 | $\sigma_r \rightarrow$ | 3.53 | 3.53±0.07 |
|  | 0.40 | 6.14781 | $\sigma_t \rightarrow$ | 4749.53 | 4749.53 |
|  | 1.5543 | 3.91659 | $\sigma_s \rightarrow$ | 3761 | 3761±6 |
| O (8,16) | 41.2421 | 5.47089 | $\sigma_r \rightarrow$ | 0.190 | 0.190±0.019 |

| | 0.40 | 5.47089 | $\sigma_t \rightarrow$ | 3761.19 | 3761.19 |
|---|---|---|---|---|---|
| | 1.02922 | 3.12533 | $\sigma_s \rightarrow$ | 1992 | 1992±6 |
| Si (14,28) | 21.3192 | 4.15456 | $\sigma_r \rightarrow$ | 177 | 177±5 |
| | 0.40 | 4.15456 | $\sigma_t \rightarrow$ | 2169 | 2169 |
| | 1.1803 | 4.03655 | $\sigma_s \rightarrow$ | 3010 | 3010±8 |
| Ca (20,40) | 39.0960 | 5.21685 | $\sigma_r \rightarrow$ | 410 | 410±20 |
| | 0.40 | 5.21685 | $\sigma_t \rightarrow$ | 3420 | 3420 |

### 8.4.3. $^{3}_{2}He$ *Cross-sections on some targets in the intermediate energy*

The cross-sections $\sigma_t$ at three different He-3 energies have been calculated for 5 different targets [$^{9}_{4}$Be, $^{12}_{6}$C, $^{16}_{8}$O, $^{28}_{14}$Si, $^{40}_{20}$Ca]. The calculated cross-sections have been compared with those measured, taken from [32, 38-40]. The comparisons are made in the way that is described above and given in table 9. There is no need for relative angular momentums here.

**Table 9.** [He [2 , 3] + Xn [Z , N] Total cross-section comparison with those measured

| Xn [Z, N] (target) | Energy (projectile) $E_L$ (MeV) | $R_0$ (fm) | $R_c$ (fm) | $R_m = r_2$ (fm) | $r_1$ (fm) (impact parameter) | Calculation $\sigma_t$ (mb) | Experiment $\sigma_t$ (mb) |
|---|---|---|---|---|---|---|---|
| Be[4,9] | 96.4 | 0.673279 | 1.40048 | 2.37051 | 2.53101 | 805 | 805±30 |
| | 137.8 | 0.623868 | 1.29770 | 2.19747 | 2.30905 | 670 | 670±30 |
| | 167.3 | 0.607056 | 1.26273 | 2.13825 | 2.23016 | 625 | 625±30 |
| C[6,12] | 96.4 | 0.620244 | 1.42000 | 2.31455 | 2.53885 | 810 | 810±40 |
| | 137.8 | 0.594922 | 1.36203 | 2.22006 | 2.37697 | 710 | 710±30 |
| | 167.3 | 0.572480 | 1.31065 | 2.13631 | 2.26556 | 645 | 645±35 |
| O[8,16] | 96.4 | 0.631332 | 1.59086 | 2.50140 | 2.78546 | 975 | 975±35 |
| | 137.8 | 0.606261 | 1.52768 | 2.40206 | 2.60079 | 850 | 850±50 |
| | 167.3 | 0.595506 | 1.50058 | 2.35945 | 2.52313 | 800 | 800±25 |
| Si[14,28] | 96.4 | 0.600731 | 1.82417 | 2.69058 | 3.15392 | 1250 | 1250±65 |
| | 137.8 | 0.603057 | 1.83124 | 2.70099 | 3.02513 | 1150 | 1150±70 |
| | 167.3 | 0.590377 | 1.79273 | 2.64420 | 2.91119 | 1065 | 1065±40 |

| Ca[20,40] | 96.4 | 0.544438 | 1.86195 | 2.64717 | 3.28976 | 1360 | 1360±90 |
| --- | --- | --- | --- | --- | --- | --- | --- |
| | 137.8 | 0.563942 | 1.92866 | 2.74200 | 3.19154 | 1280 | 1280±85 |
| | 167.3 | 0.565988 | 1.93565 | 2.75195 | 3.12222 | 1225 | 1225±75 |

The calculation of cross sections through solution of radial Schrödinger equation by the partial wave expansion is very difficult. In many cases, some approximations are needed for these kinds of solutions. In the present study, firstly, differential elastic scattering, inelastic (or reaction) scattering and total cross-sections have been calculated without using any approximation. These calculations have been performed using a simple method, improved for the solution of RSE, for an incident particle being in a central field of any form. We have obtained the general formulas of the scattering amplitudes and elastic, inelastic (no-elastic) and total scattering cross-sections. Secondly, we have made some applications. In these applications, the potentials have been assumed to have Saxon-Woods shape plus spin-orbit interaction, centrifugal and Coulomb potentials. With these potentials, first, for the thermal neutrons; the elastic, inelastic (neutron radiative capture) and total scattering cross-sections of different targets have been calculated. Later, total scattering cross-sections for $^3_2$He particles of three different energies on 5 targets have been calculated. The calculated results have been compared with experimental results. The results calculated have given satisfactory agreement with the available experimental results. It can be found more cross section calculations in the reference [33].

The calculations have also shown that the total cross sections depend on the mean potential range. Thus, it is also proved that the total cross-sections can be calculated easily using even very complex potentials. The same calculations have also been performed using optical potentials, but the results have not been included here due to getting the same scattering distance and cross-sections. The use of two parameters is seen to be enough in the agreement of the calculated results with those measured results, whereas this agreement is ensured using more parameters in the partial wave expansion method.

## 9. Conclusion and some explanations

We have found a simple procedure for the general solution of the time-independent SE in one dimension without using any infinite series or other approximations. The wave functions, which are always periodic at the bound states, are given by the equations (4.1 - 4.5). In our procedure of solution, there are two difficulties: one is to solve the equation $E = U(x)$ to find the energy; and the other is to integrate $\sqrt{U(x)}$, namely, to calculate $\int U(x)dx$ to find the exact normalized wave function. If these calculations cannot be done analytically then it should be done by numerical methods. To find the energy values there is no need to calculate this integral, it is sufficient to find the classical turning points by solving the equation $E = U(x)$. Thus there is no need to know the wave functions to find the energy values of the states; it is enough to know only the potential energy function. SE has been solved for a particle in many potential wells and found their energies and normalized wave functions as examples.

Using simple procedure, the solution of the radial SE for spherically symmetric potentials without using any infinite series has been found in this study. The wave functions which are always periodicals are given by the Equations (4.1-4.5). By using this procedure, the radial SE has been solved for a particle found in many spherically symmetric central potential wells and two different solutions have been found. One of them is symmetric function and the other is antisymmetric function. From these expressions, it is observed that these functions are periodic and they are similar to each other in form for all potential wells. This simple solution was applied to scattering and tunneling theories and it was seen to give good results.

The solution that we propose is a general solution. The points of view supporting the method I present here is more realistic which are as follows:

As it is known, SE is a second order differential equation with variable coefficients. The solutions of such equations are based generally on series method and special functions (Hermit, Bessel …) in quantum physics. In the expanding of power series, the consecutive relations between the coefficients of the series are found. Some approaches are taken to make the series convergent and by using them, the energy values and the wave functions are determined. In these solutions, one or more approximations are used. Some other methods such as perturbation, WKB, variation, etc. are also applied for solving the SE and approximate solutions are obtained. However, the solution that is proposed here is neither based on series methods nor special functions, and no approximation is used. I think that these solutions, which do not have any approach, are more realistic.

It is seen that the functions found with the known methods and with the present method are different in form. However, when we do numerical calculations, the results are not very different from each other and are consistent with their well-known values. For example the one dimensional harmonic oscillator, our result and the result obtained by the known methods are very close to each other.

SE, the fundamental equation of the quantum mechanics, is also known as the Schrödinger wave equation. In physics, the harmonic waves are represented with periodic functions e.g. sin and cos. However, most of the known solutions of the SE are in the form of polynomials. Especially the solution of the harmonic oscillator should be periodic function. However, the known solution is polynomial. As it is seen in the functions (5.1/3), our solution gives sin and cos functions, which makes it more realistic.

It is said that the quantum mechanics includes classical mechanics. Hence, the results of the classical mechanics should be obtained from the results found in quantum mechanics. As it is seen from the functions given in (5.1/4), it is very difficult to obtain **sin** and **cos** functions from polynomials. However, the classical solutions could be easily obtained if $\Psi(x)$ is taken very small in (5.1/4) functions.

In quantum mechanics textbooks, it is said that all infinitely high potential wells are similar to each other. But when we look at their solutions, we see that all of the solutions are different in form. On the contrary our solutions for this kind of well are similar to each other in form.

In the quantum mechanics, if there is no exact solution, sometimes, variation principle is used to find the ground state energy. In the calculations of variation, the calculations are made with a trial function, and it is seen that the results are not very dependent on the trial function. However the wave functions are not necessary to find the ground or excited state energy in our solutions. It is sufficient to know the classical turning points of the potential function.

Special conditions are not required in order to find energy values and the wave functions. The continuity of the wave functions and their derivatives at the classical turning points are enough.

This paper can be considered as a summary of our several previous articles. The application of our procedure is very easy. Most of the problems that could not be solved analytically with the known methods can easily be solved using our procedure. A complete solution of the SE used in all branches of physics has been made. A problem that has been worked by many theoretical physicists has been solved. I think that this solution is very useful and helpful for those interested physicists in quantum mechanics and applications.

**Acknowledgements**
I would like to express my sincere gratitude to my wife Özel and my daughters Işıl and Beril for their support and patience during my study and for their help in editing.